\documentclass[twocolumn]{aa} % for a paper on 1 column  
\usepackage[varg]{txfonts}
\usepackage{graphicx}
\usepackage{txfonts}
\usepackage{multirow}
\usepackage{xcolor}
\usepackage{tabularx}
\usepackage{array}
\usepackage{rotating}
\usepackage{makecell}
\usepackage{tablefootnote}
\usepackage{booktabs}
\usepackage{threeparttable}
\usepackage{siunitx}
\usepackage{soul}
\usepackage[utf8]{inputenc}
\usepackage{hyperref}
\def\ew{$W_{2796}$}

\def\kms{km~s$^{-1}$}

\def\mgii{Mg~{\sc ii}~} 
\def\oii{O~{[\sc ii}]~} 
\def\oiii{O~{[\sc iii}]~}

\def\oiiiab{[O~{\sc iii}] $\lambda\lambda$4959,5007} 
 
\def\oiiib{[O~{\sc iii}] $\lambda$5007~}

\def\oiii{[O~{\sc iii}]} 
\def\oiiab{[O~{\sc ii}] $\lambda\lambda$3727,3729}  
  
\def\oii{[O~{\sc ii}]}

\def\hbeta{H$\beta$~}

\DeclareUnicodeCharacter{02BC}{'}
\def\sample{606}

\def\samplesightlinesvhs{185}
\def\samplesightlinesviking{100}

\def\samplesnroiigetwopfive{603}
\def\samplesnroiigethree{578}
\def\samplesnroiigetwopfiveoiiiandhbeta{53}
\def\samplesnroiigetwopfiveonlyoiii{85}
\def\samplesnrhboiiigethree{three}
\def\samplethreefilterdetection{508}
\def\zabsmin{0.3}
\def\zabsmax{1.6}
\def\detection{377}
\def\detectionrate{0.45}
\def\detectionewgetwo{277}
\def\detectionewgetworate{73.5}
\def\detectionewgeonerate{95.8}
\def\mstarmin{8.4}
\def\mstarmax{11.6}
\def\logsfrmin{-1.2}
\def\logsfrmax{2.7}
\def\ipmin{4}
\def\ipmax{24}

\def\detectionsec{229}
\def\bmgiizabsmin{0.4}
\def\bmgiizabsmax{1}
\def\bmgiidetection{216}
\def\bmgiilogsfrmin{-3.7}
\def\bmgiilogsfrmax{2.3}
\def\bmgiimstarmin{7.9}
\def\bmgiimstarmax{11.1}
\def\primarydetection{593}
\def\primaryipmin{4}
\def\primaryipmax{24}
\def\primaryzabsmin{0.3}
\def\primaryzabsmedian{0.8}
\def\primaryzabsmax{1.6}

\def\primarymstarmin{7.9}
\def\primarymstarmedian{9.9}
\def\primarymstarmax{11.6}
\def\primarylogsfrmin{-1.9}
\def\primarylogsfrmedian{0.9}
\def\primarylogsfrmax{2.7}
\def\primarystarburstrate{13}

\def\primarysfroiimin{0.1}
\def\primarysfroiimax{14.9}
\def\logcharacteristicoiilum{40.59}
\def\primarylumoiinormmin{0.2}
\def\primarylumoiinormmax{68.1}
\def\primaryhighsfrdensity{458}
\def\primaryhighsfrdensityrate{77.2}
\def\primarylowzabsmean{0.4}
\def\primarylowzabsewmean{2.1}

\def\primarymidzabsewmean{2.4}

\def\primaryhighzabsmean{1.2}
\def\primaryhighzabsewmean{2.9}

\def\primarylepthirtyzabs{178}
\def\primarylepthirtyzabsmean{0.5}

\def\primarygepseventyzabsmean{1.2}
\def\primarylowzabs{52}
\def\primarymidzabs{396}
\def\primaryhighzabs{145}
\def\primarygepszabsalphaerrlow{0.06}
\def\primarygepszabsalpha{0.45}
\def\primarygepszabsalphaerrhigh{0.06}
\def\primarygepszabsbetaerrlow{0.46}
\def\primarygepszabsbeta{-0.12}
\def\primarygepszabsbetaerrhigh{0.42}
\def\primaryleptzabsalphaerrlow{0.05}
\def\primaryleptzabsalpha{0.32}
\def\primaryleptzabsalphaerrhigh{0.05}
\def\primaryleptzabsbetaerrlow{0.60}
\def\primaryleptzabsbeta{0.30}
\def\primaryleptzabsbetaerrhigh{0.63}
\def\leptzabshm{69}
\def\gepszabshm{134}
\def\lzhzmassstat{0.11}
\def\lzhzmassstatpvalue{0.54}
\def\lzhmsfrmedian{3.6}
\def\hzhmsfrmedian{21.4}

\def\lzhmewmean{2.1}
\def\lzhmewstd{0.1}
\def\hzhmewmean{2.9}
\def\hzhmewstd{0.1}
\def\lzhzmgiiewstat{0.39}
\def\lzhzmgiiewstatpvalue{0.001}
\def\rttlzzabsmedian{0.5}
\def\rtthzzabsmedian{0.9}
\def\rttlmmstarmin{7.9}
\def\rttlmmstarmedian{9.5}
\def\rttlmmstarmax{9.8}
\def\rtthmmstarmin{9.8}
\def\rtthmmstarmedian{10.1}
\def\rtthmmstarmax{11.1}
\def\primarydvmin{-389}
\def\primarydvmean{-24}
\def\primarydvmax{364}
\def\primarydvstd{88}
\def\primarydvlm{101}
\def\primarydvmm{312}
\def\primarydvhm{176}
\def\primarylmdvmean{-11}
\def\primarymmdvmean{-26}
\def\primaryhmdvmean{-28}
\def\primarylmdvsigma{73}
\def\primarymmdvsigma{86}
\def\primaryhmdvsigma{96}
\def\primarylmdvsigmamin{58}
\def\primarylmdvsigmamax{83}
\def\primaryhmdvsigmamin{79}
\def\primaryhmdvsigmamax{106}

\begin{document} 

   \title{Baryonic Ecosystem IN Galaxies (BEINGMgII) - III. Cool gas reservoirs at $0.3 \le z \le 1.6$ in the Dark Energy Survey}
   %\subtitle{I. Overviewing the $\kappa$-mechanism}
   \author{ 
        Reena Chaudhary\inst{1,2}, %\fnmsep\thanks{},
        Ravi Joshi\inst{1},
        Sarbeswar Das\inst{1}, 
        Michele Fumagalli\inst{3,4},
        Glenn G. Kacprzak\inst{5},
        Matteo Fossati\inst{3,6}, 
        Celine P\'eroux\inst{7,8},
        \and
        Luis C. Ho\inst{9,10}
    }
   \institute{Indian Institute of Astrophysics (IIA), Koramangala, Bangalore 560034, India \\     \email{rvjoshirv@gmail.com}
   \and
            Pondicherry University, R.V. Nagar, Kalapet, Puducherry 605014, India
            \and
             Universit\'a degli Studi di Milano-Bicocca, Dip. di Fisica G. Occhialini, Piazza della Scienza 3, 20126 Milano, Italy
             \and
             INAF - Osservatorio Astronomico di Trieste, via G.B. Tiepolo 11, I-34143 Trieste, Italy
             \and 
             Centre for Astrophysics and Supercomputing, Swinburne University of Technology, Hawthorn, VIC 3122, Australia
             \and    
             INAF - Osservatorio Astronomico di Brera, Via Brera 28, 21021, Milano, Italy
             \and
             European Southern Observatory, Karl-Schwarzschildstrasse 2, D-85748 Garching bei Munchen, Germany
             \and 
             Aix Marseille Universit\'e, CNRS, LAM (Laboratoire d’Astrophysique de Marseille) UMR 7326, F-13388 Marseille, France
            %  Key Laboratory for Research in Galaxies and Cosmology, Shanghai Astronomical Observatory, Chinese Academy of Sciences, 80 Nandan Road, Shanghai 200030, People’s Republic of China
            % \and 
            % Kavli Institute for the Physics and Mathematics of the Universe (WPI), The University of Tokyo Institutes for Advanced Study (UTIAS), The University of Tokyo, 5-1-5 Kashiwanoha, Kashiwa-shi, Chiba 277-8583, Japan
             % \and 
             % Department of Astronomy, University of Virginia, 530 McCormick Road, Charlottesville, VA 22904,USA
             \and 
             Kavli Institute for Astronomy and Astrophysics, Peking University, Beijing 100871, Peopleʼs Republic of China
             \and
             Department of Astronomy, School of Physics, Peking University, Beijing 100871, Peopleʼs Republic of China
             }
   \date{Received July 26, 2024; accepted July 26, 2024}

% \abstract{}{}{}{}{} 
% 5 {} token are mandatory
 \titlerunning{BEINGMgII - Nature of galaxies harbouring cool gas reservoirs}
  \authorrunning{Chaudhary et al.}
  \abstract
  % context heading (optional)
  % {} leave it empty if necessary  
    {We investigate the origin of intervening cool \mgii absorption detected in the spectra of background quasars and the nature of associated galaxies across a broad redshift range of $\zabsmin \le z \le \zabsmax$. Using nebular \oiiab\ emission lines identified in DESI fiber spectra centered on quasar, we detect \detection\ galaxies at a typical detection rate of $\sim$\detectionrate\% at $z \lesssim 1$, which increases with \mgii equivalent width (\ew). A significant fraction (\detectionewgetworate\%) of these galaxies are associated with strong absorbers with \ew\ $\ge$ 2~\AA. These absorbers trace galaxies spanning stellar masses of $\rm \mstarmin \le \log(M_{\star}/M_{\odot}) \le \mstarmax$ and star formation rates of $\rm \logsfrmin \le \log(SFR~[M_{\odot}yr^{-1}]) \le \logsfrmax$, located at projected galactocentric distances of \ipmin–\ipmax\ kpc. We find the average \mgii absorber strength increases from \primarylowzabsewmean~\AA\ to \primaryhighzabsewmean~\AA\ between redshifts $z \sim$ \primarylowzabsmean\ and \primaryhighzabsmean, indicating evolution in the cool gas content of galaxy halos. The relatively constant absorber strength with galactocentric distance implies a clumpy structure of cool gas in the circumgalactic medium (CGM). Further, we find a positive correlation between \ew\ versus stellar mass ($M_\star$), and star formation rate (SFR), suggesting that the distribution of metal-enriched cool gas in the CGM is closely tied to the properties of the host galaxies. The redshift evolution of gas-phase metallicity suggests that strong \mgii absorbers trace the general population of star-forming galaxies. The velocity dispersion of the cool gas increases with halo mass, and the wide range of line of sight velocity offset (\primarydvmin\ to \primarydvmax ~\kms) between the galaxy systemic velocity and absorbers    highlights the dynamical nature of CGM. However, the majority of this gas remains gravitationally bound to the dark matter halos, consistent with a picture of gas recycling via galactic fountains.}
  \maketitle
%
%-------------------------------------------------------------------

\section{Introduction}
\label{sec:intro}

Galaxies across the wide spectrum of stellar mass and morphologies, including actively star-forming and quiescent galaxies, harbor cool gas reservoirs within their halo \citep{Chen2010ApJ...714.1521C,Nielsen2013ApJ...776..115N,Huang2021MNRAS.502.4743H,Chen2025ApJ...981...81C}. This metal-rich cool gas extending up to the virial radius is believed to originate in galactic winds driven by the AGN feedback or supernova explosions \citep{Rahmati2015MNRAS.452.2034R, Schroetter2019MNRAS.490.4368S,Ho2021ApJ...923..137H,Huscher2021MNRAS.500.1476H,Zabl2021MNRAS.507.4294Z,Schroetter2024A&A...687A..39S}. The supernova-driven outflow models require high-energy coupling from the supernovae to the gas, and with supernova efficiencies of nearly unity in transferring energy to the wind~\citep{Afruni2021MNRAS.501.5575A}. Furthermore, the continuous accretion of pristine gas from the
intergalactic medium replenishes the gas reservoir, the circumgalactic medium (CGM), which is fundamental for maintaining the galaxy growth \citep{Tumlinson2017ARA&A..55..389T}.  However, tracing this low-density gas remained challenging \citep{Peroux2020MNRAS.499.2462P,Peroux2024arXiv241107988P}.

The \mgii absorption in the background quasar spectra gives access to the cool photoionized gas of temperature $\simeq{10^4}$K and high neutral hydrogen column density of log$\rm N (H${~\sc i}$) [\rm cm^{-2}] \sim 18-22$ \citep{Tumlinson2017ARA&A..55..389T, Peroux2020MNRAS.499.2462P}. The studies based on absorber-galaxy connection show that \mgii absorbers are generic feature near ordinary galaxies, largely traces the underlying gas kinematics around them, including inflowing and outflowing gas \citep{Bouche2007ApJ...669L...5B, Nestor2011MNRAS.412.1559N, Ho2017ApJ...835..267H,Lan2018ApJ...866...36L, Zabl2021MNRAS.507.4294Z, Guo2023Natur.624...53G, Bacon2023A&A...670A...4B, Nateghi2024MNRAS.534..930N,Kacprzak2025arXiv250711613K};  the multiple halos of a galaxy group \citep{Kacprzak2010MNRAS.406..445K,Bielby2017MNRAS.468.1373B, Peroux2017MNRAS.464.2053P,Nielsen2018ApJ...869..153N,Fossati2019MNRAS.484.2212F,Dutta2020MNRAS.499.5022D, Lundgren2021ApJ...913...50L,Antonia2024MNRAS.531.3658F}; the intra-group medium \citep{Gauthier2013MNRAS.432.1444G}; and/or cool stripped gas from environmental processes \cite{Dutta2020MNRAS.499.5022D}. The statistical analysis of millions of quasar-galaxy pairs revealed an excess mean \mgii absorption extending on a spatial scale of 20 kpc to 10 Mpc, with an excess around emission line galaxies (ELGs) 
than the luminous red galaxies \citep{Lan2018ApJ...866...36L, Wu2025ApJ...983..186W, Chen2025ApJ...981...81C}. The studies based on cosmological hydrodynamic zoom-in simulations further highlight that the CGM around L* star-forming galaxies has experienced substantial evolution in both its physical structure and dynamical state from the present epoch until cosmic noon, i.e., $z \sim 2-3$ \citep{Huscher2021MNRAS.500.1476H}. The signature of cool gas evolution is further supported by the stronger \mgii absorption in halos of high-$z$ galaxies than the low-$z$ counterparts \citep{Wu2025ApJ...983..186W}. On the contrary, \citet[][]{Chen2012MNRAS.427.1238C} found a similar spatial extent and mean absorption equivalent width in the CGM at low and high redshifts across the 11 Gyr \citep[see also,][]{Rudie2019ApJ...885...61R}.

The strong interplay between galaxy and CGM is further highlighted by the observed increase in absorption strength with the SFR and stellar mass of absorber host galaxies,  supporting the galactic wind origin of \mgii gas \citep{Bouche2007ApJ...669L...5B, Zibetti2007ApJ...658..161Z,Bouche2012MNRAS.426..801B, Zabl2019MNRAS.485.1961Z, Schroetter2019MNRAS.490.4368S}. Mapping the disk–halo interface within $\lesssim 50$ kpc, where gas accretion and outflows are more pronounced, offers a pathway to understand how the physical state of the CGM regulates feedback, accretion, and long-term galaxy growth \citep{Bouche2007ApJ...669L...5B}. However, finding quasar-galaxy pairs at these close separations remained observationally expensive. In previous studies,  the absorber host galaxies are generally found at an average galactocentric distance of $\sim$100kpc \citep{Kacprzak2008AJ....135..922K,Chen2010ApJ...714.1521C, Nielsen2013ApJ...776..115N, Lundgren2021ApJ...913...50L}. In MusE GAs FLOw and Wind (MEGAFLOW) studies, targeting the CGM at close galactocentric distances ($\lesssim 100$ kpc) around galaxies at intermediate redshifts ($z \sim 0.5$–$1.5$), found the signature of biconical outflow and co-rotating gaseous discs, suggesting that the mass accretion rates from the CGM are adequate for sustaining star formation \citep{Schroetter2019MNRAS.490.4368S, Zabl2021MNRAS.507.4294Z,Schroetter2024A&A...687A..39S}.  \par

The fiber-based spectroscopic surveys, using a finite fiber size of radius $1-1.5$ arcsec, trace a physical area of 5-10 kpc over a redshift window of $0.3 \le z  \le 1.5$ around the central quasar. Thus, proven to be a powerful probes of galaxies at quasar proximity of $\lesssim 20$~kpc \citep{Noterdaeme2010MNRAS.403..906N,Straka2015MNRAS.447.3856S,Joshi2017MNRAS.471.1910J,Das2025A&A...695A.207D}. \citet[][hereinafter Paper II]{Das2025A&A...695A.207D}, investigated the galaxies hosting \mgii absorbers over a redshift range of $\bmgiizabsmin \le z \le \bmgiizabsmax$ by searching strong nebular emission lines (\oii, \oiii, H$_{\beta}$) in background quasar spectra and associated photometric counterparts within 2 arcsec of quasar proximity in DECaLS imaging survey. It resulted in \bmgiidetection\ elusive quasar-galaxy pairs at exclusively low impact parameters of $\lesssim 20$~kpc, predominantly tracing galaxies with star formation rate (SFR) of $ \bmgiilogsfrmin \le \rm \log SFR [M_{\odot}\ yr^{-1}] \le \bmgiilogsfrmax$ and stellar mass ranging between \bmgiimstarmin $\le \rm log(M_{\star}/M_{\odot}) \le $ \bmgiimstarmax. A positive correlation between the absorber strength and SFR and specific SFR, supports the wind origin. It is further reinforced by the near-constant absorption intensity at low-impact parameters, which points to the possible contribution of strong outflows and satellite galaxies in the metal enrichment of the galaxy halo \citep{Weng2024MNRAS.527.3494W}.

In this study, we advance the investigation of \mgii absorber host galaxy from Paper II by leveraging 270,529 \mgii absorbers from the value-added catalog of the Dark Energy Spectroscopic Instrument (DESI) Data Release-1 \citep{Napolitano2023AJ....166...99N} and $\approx$ 14,000 square degrees imaging data from the Dark Energy Spectroscopic Instrument (DESI) Legacy Surveys (DECaLS\footnote{\href{https://www.legacysurvey.org/decamls/}{https://www.legacysurvey.org/decamls/}}) \citep{Dey2019AJ....157..168D}. This paper is structured as follows. Section 2 describes our sample selection and analysis. Section 3 examines the absorber galaxy association and the nature of galaxies associated with \mgii absorbers. The conclusions of this study are summarized in Section 4. Throughout, we have assumed a flat Universe with $H_0$ = 70 $\rm km\ s^{-1}\ Mpc^{-1}$, $\Omega_m$ = 0.3, and $\Omega_\Lambda$= 0.7.

\begin{figure*}
    \centering
    \includegraphics[width=0.95\textwidth]{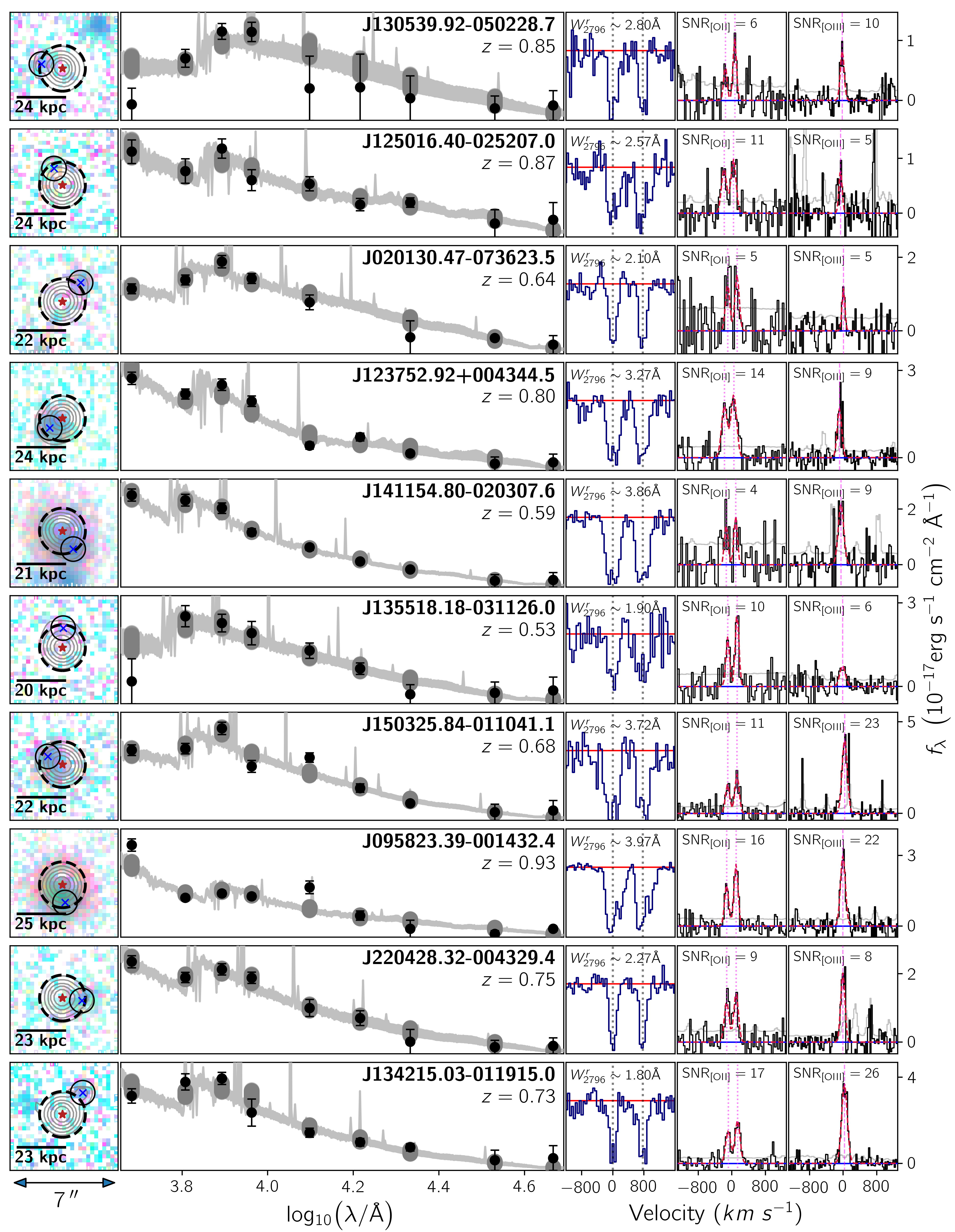}
    \caption{Postage stamps DECaLS colour composite, centred on the quasar. The SDSS fibre, with a radius of \SI{1.5}{\arcsecond}, is indicated by dashed circles, while the solid circular aperture highlights the MgII host galaxy. The second column exhibits the multi-band SED fit at the absorber redshift. Columns three, four, and five show the detected \mgii absorption, \oiiab\ and \oiiib\ nebular emission line from the \mgii absorber host, respectively.}
    \label{fig:collage}
\end{figure*}

\section{Sample and Analysis}
We use the DESI-DR1 value-added catalog of \mgii absorbers, comprising 137,328 quasar sightlines, hosting 162,836 absorbers with redshift, $z_{abs} \le 1.6$, ensuring a coverage of  \oiiab\ nebular emission in DESI spectra. To find the absorber host galaxy, we follow the procedure adopted in  Papar II. In brief, firstly, we search the quasar sightlines having at least one photometric counterpart within 2 arcsec around the quasar in the DECaLS imaging survey \footnote{\href{https://datalab.noirlab.edu/data/legacy-surveys}{https://datalab.noirlab.edu/data/legacy-surveys}}. The DECaLS covers $\approx 14000$ sq. degrees of sky in $g, r, z$ optical passbands at a $5\sigma$ survey depth of $g=24.0$, $r=23.4$, and $z=22.5$ AB magnitudes \citep{Dey2019AJ....157..168D}.  In this work, we extend the \mgii absorber host galaxy search beyond the redshift window of $z \sim 1$   adopted in Paper II, and analyze 52,369 SDSS quasar sightlines hosting 58,993 \mgii absorbers within $1 \leq z \leq 1.5$ \citep{Zhu2013ApJ...770..130Z,Anand2021MNRAS.504...65A}. The sample summary is given in Table~\ref{tab:sample}, including 24848 and 9193 photometric objects near quasars from DESI and SDSS, respectively.

\begin{table}[htbp]
\centering
\caption{Summary of the sample of \mgii absorbers}
\label{tab:sample}
{\scriptsize
\begin{tabular}{p{0.8cm}|p{0.8cm}|p{0.7cm}|p{0.8cm}|p{1.4cm}|c}

\toprule
Survey & \mgii     &    Galaxies            & \oii           & No. of bands   & SED  \\
       & absorbers &   $<2''$               & emission       & with $4\sigma$ &   \\
       &           &                        &                & detections     & 
\\ 
% \cmidrule(lr){5-7}
%      &        & Absorbers      &  $<2''$       & [O\,\textsc{ii}]& H$\beta$ & [O\,\textsc{iii}]  & Detection  \\ 
\midrule
\multirow{3}{*}{DESI} & \multirow{3}{*}{162,836} & \multirow{3}{*}{24,848} & \multirow{3}{*}{522} & \multirow{2}{1cm}{$\ge3$ (440) } & $341^a$  \\
\cline{6-6}
                      &                          &                         &                      &                      & $99^b$  \\
\cline{5-6}
                      &                          &                         &                      &         $<3$ ($82^b$)           &         -          \\
\cline{1-6}
\multirow{3}{*}{SDSS} & \multirow{3}{*}{58,993}  & \multirow{3}{*}{9,193}  & \multirow{3}{*}{84}  & \multirow{2}{1cm}{$\ge3$ (68)}  & $36^a$  \\
\cline{6-6}
                      &                          &                         &                      &                      & $32^b$  \\
\cline{5-6}
                      &                          &                         &                      &          $<3$ ($16^b$)          &         -             \\
% Add more rows as needed
\bottomrule
\end{tabular}
\footnotesize{
\begin{flushleft}
a: Primary Sample; b: Secondary Sample\\
\end{flushleft}}}
\end{table}

\begin{figure*}
    \centering
    \includegraphics[width=0.95\textwidth]{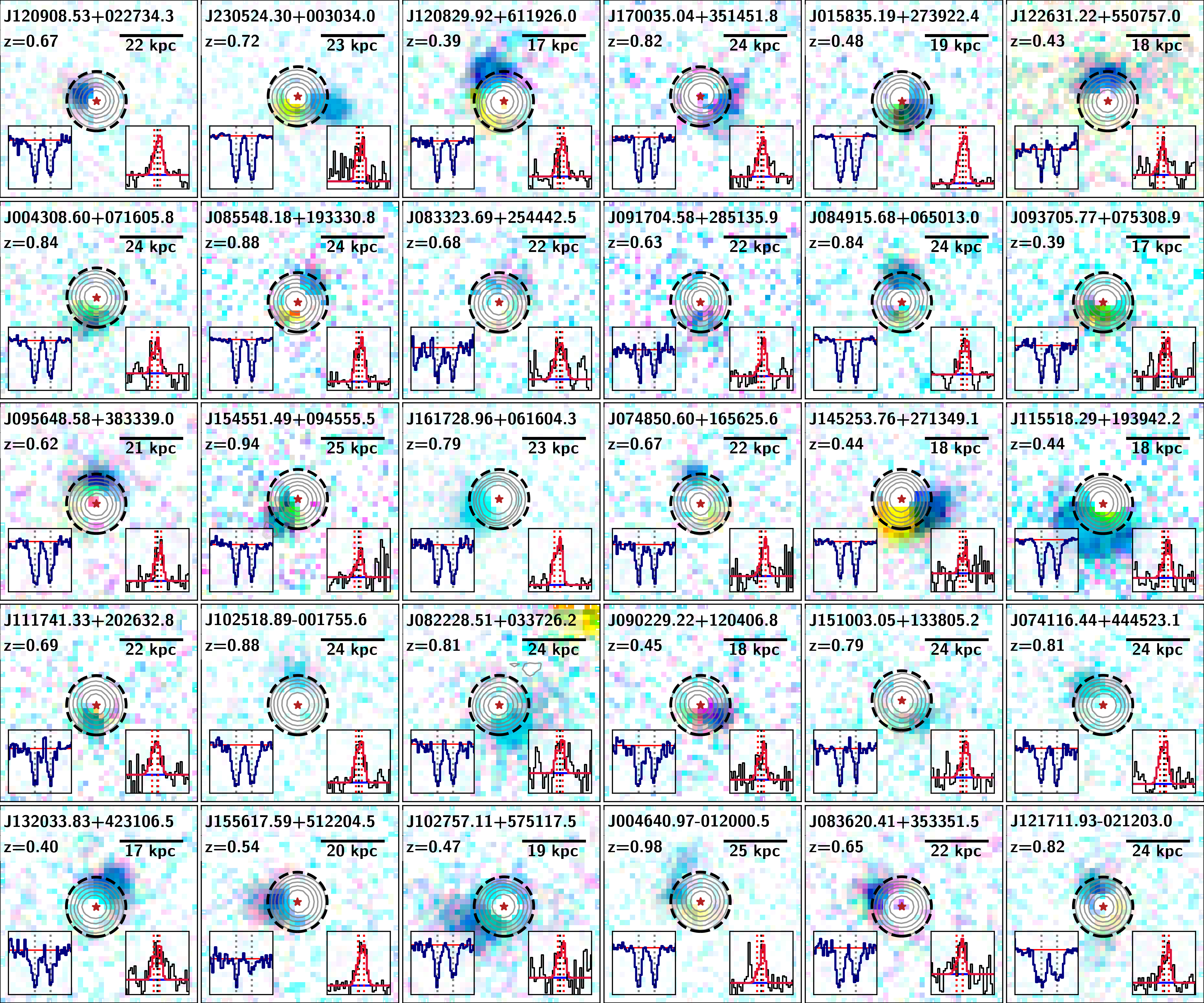}
    \caption{ The faint \mgii absorber galaxy revealed in the DECaLS residual images. These galaxies remained undetected in DECaLS photometry due to the glare of the bright quasar. Each panel shows the $g$ (blue), $r$ (green),  $z$ (red) band colour composite inverted-image centered on the quasar (grey contour), marked as {\it star}, and the  SDSS fibre of radius 1.5 arcsec ({\it dotted circle}). The associated \mgii absorption and {\oiiab} nebular emission line profiles are shown in the bottom left and right insets, respectively.}
    \label{fig:residual}
\end{figure*}

\subsection{Searching nebular emission lines}
Next, for the absorber-galaxy association, we search for the strong nebular emission lines from potential host galaxies. Following Paper II, we searched for the \oiiab\ nebular emission for each \mgii absorber close to the absorber redshift ($z_{\rm abs}$) in the quasar continuum-subtracted spectrum. We model the emission feature with a Gaussian kernel over the velocity range of $\pm 400 \rm km s^{-1}$ around $z_{\rm abs}$  and determined the detection significance based on the signal-to-noise ratio defined in \citet{Bolton2004AJ....127.1860B} as $S/N = \frac{\sum_{i}^{} f_i u_i / \sigma_i^2}{\sqrt{\sum_{i}^{} u_i^2/\sigma_i^2}}$, where $f_i$ is the residual line flux in the $i^{th}$ pixel, $\sigma_i$ is the flux error, and $u_i$ is a Gaussian kernel, normalized such that $\sum_i u_i = 1$. The position and line width of the kernel were determined by minimizing the $\chi^2$ over the defined velocity window. This resulted in a sample of \samplesnroiigetwopfive\ galaxies with \oii\ emission line detected at $> 2.5\sigma$ level, including \samplesnroiigethree galaxies at \oii  $> 3\sigma$ level. Furthermore, we searched H$\beta$, and \oiiiab\ nebular emissions from the galaxy. We performed a visual scrutiny for any bad pixels, continuum, or strong skyline regions.  Among \samplesnroiigetwopfive\ \oii\ selected galaxies, we found \samplesnroiigetwopfiveoiiiandhbeta\ galaxies with additional \oiiib\ and H$\beta$  as well as \samplesnroiigetwopfiveonlyoiii\ galaxies with only \oiiib\ nebular emission, detected at $\ge 3\sigma$ level. Further,  \samplesnrhboiiigethree\ additional systems were found based on only \oiiib\ and H$\beta$ nebular emission, resulted in a final sample of  \sample\ \mgii galaxies.
The sample summary, including the absorber sightlines, proximate galaxies, and emission line-selected galaxies, is given in Table~\ref{tab:sample}.

\subsection{\mgii absorber host in DECaLS residual images} 
To search for galaxies producing strong \mgii absorption, we further explore the residual images provided by DECaLS. For this, we explore DESI residual images with non-zero residual flux at the quasar location with 5 contiguous pixels at $\ge 3\sigma$ level. The residual images are obtained using the Tractor algorithm, which models each source, detected in three individual-band image stacks at the 6$\sigma$ level. In the Tractor, each source is modelled using a small set of parametric light profiles, including a delta function for point sources, a de Vaucouleurs law, an exponential disc, or a composite de Vaucouleurs plus an exponential. Assuming the same model across all the bands, the same light profile (an exponential disc, de Vaucouleurs, or combination) is consistently fit to all images to determine the best-fit source position, source shape parameters, and photometry \citep{Dey2019AJ....157..168D}. We note that modeling uncertainties related to source morphology, blending, and saturation may lead to more false positives in residual-based searches. To mitigate this, we firstly searched for the \oii\ nebular emission for all \mgii absorbers in our sample (see column 1 Table~\ref{tab:sample}). Since our search relies solely on \oii\ emission, we impose a strict detection criterion of $\ge 3\sigma$ to reduce false positives, which resulted in 900 detections \citep[see also,][]{Joshi2017MNRAS.471.1910J}. Among them, we found 562 bona fide \mgii absorber host galaxies with \oii\ nebular emission and photometric counterpart in the residual images.  Figure~\ref{fig:residual} shows examples of absorber galaxy candidates identified in DECaLS residual images.

\subsection{Modelling the Spectral Energy Distribution}
\label{sec:nebularemi}
To determine the key physical parameters of a galaxy, such as star-formation rate and stellar mass, we performed multi-band SED fitting.  From the DECaLS photometric catalog, we include DECaLS optical passband $g, r, i, z$ fluxes,  along with the mid-infrared passbands $W1$ and $W2$ from unblurred Wide-field Infrared Survey Explorer (unWISE) co-added images reaching up to a $5\sigma$ depth of $20.0$ and $19.3$ AB magnitudes in $W1$ and $W2$, respectively  \citep{Dey2019AJ....157..168D}. The mid-infrared unWISE fluxes were obtained 
with forced photometry where each optically detected source is modeled by forcing the location and shape of the model by convolving with the WISE PSF and fitting to the WISE stacked image \citep[see also,][]{Lang2016AJ....151...36L}. Furthermore, we searched for near-infrared $J, H, Ks$ passband images from the VISTA Surveys\footnote{\href{http://vsa.roe.ac.uk/}{http://vsa.roe.ac.uk/}}. We find  \samplesightlinesvhs\ sources in the VHS (at a 5$\sigma$ depth of J= 21.2, H=20.6, Ks=20.0) survey and \samplesightlinesviking\ in the VIKING (at a 5$\sigma$ depth of J= 22.1) survey.  We obtain the near-IR fluxes by simultaneously modelling the quasar and the galaxy at the centroid obtained from DESI images. In the case of non-detection, we have estimated the flux upper limits at a 4$\sigma$ level. For reasonable SED model parameters, we demanded that the target be detected (at $4\sigma$ level) in any of the three filters mentioned above, resulting in \samplethreefilterdetection\ sources, of which more than $\sim$75\% of the systems have a minimum four-band detection. The galaxy SED were modelled using {\sc BAGPIPES}\footnote{\href{https://bagpipes.readthedocs.io/en/latest/}{https://bagpipes.readthedocs.io/en/latest/}} \citep{Carnall2018MNRAS.480.4379C} and the MultiNest sampling algorithm \citep{Feroz2009MNRAS.398.1601F}. For this, we fixed the galaxy redshift determined based on \oii\ nebular emission. Here, we used a simple model considering a delayed star formation history with a wide parameter space for the age, between 50 Myr to 13.5 Gyr, mass formed ( $ 6 \le \rm log(M_*/M_{\odot}) \le 13$), and metallicity ($\rm 0.005 < [Z/H] < 5$). We assumed the dust extinction law of \citet{Calzetti1994ApJ...429..582C} with a total extinction of $0 < A_v < 4$. \par

\begin{figure}
    \centering
    \includegraphics[width=0.48\textwidth]{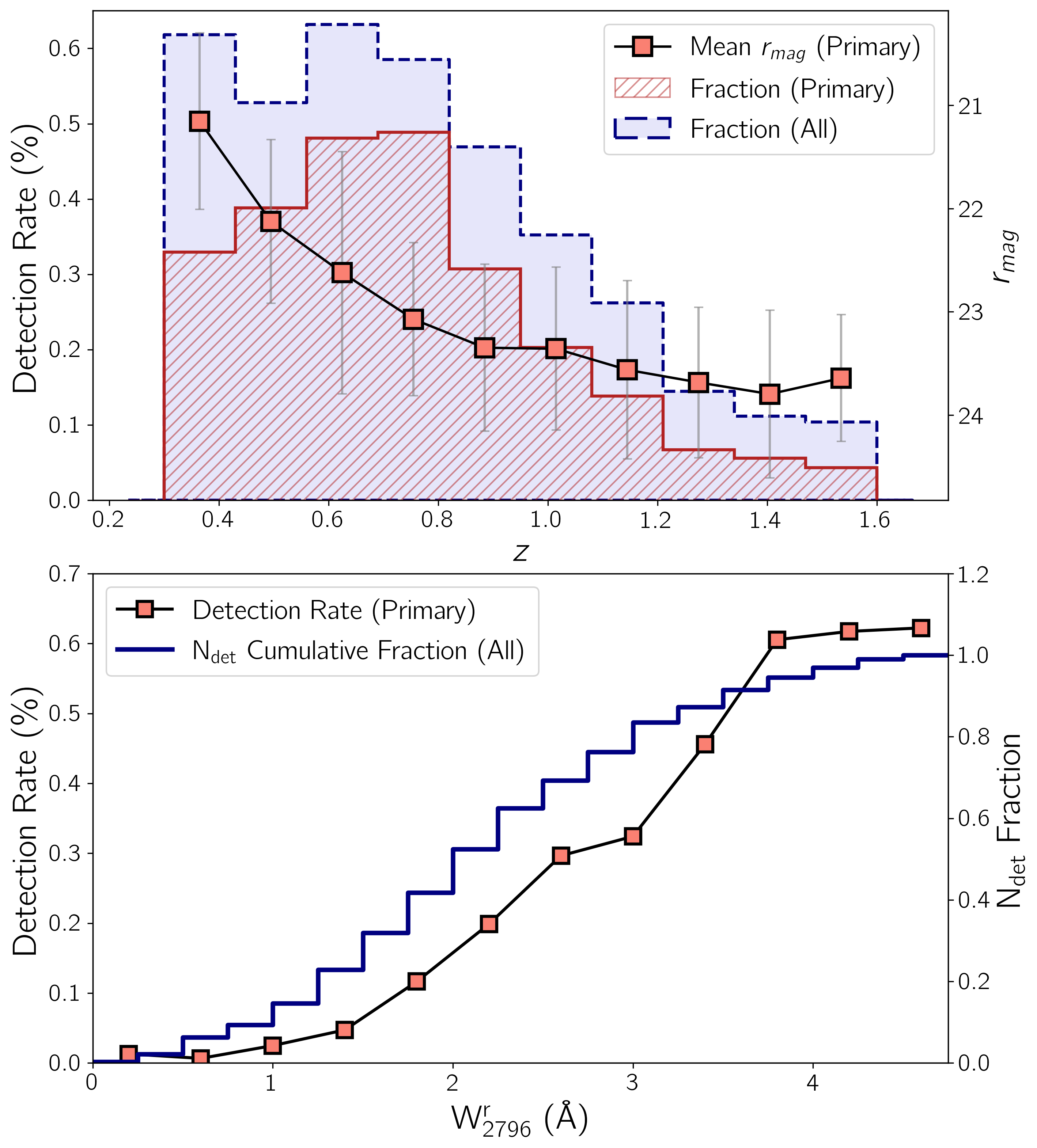}
    \caption{ Detection probability of \mgii absorbers. {\it Top panel:} Detection fraction as a function of redshift for primary (hatched), and after incorporating the secondary (dashed) set. The average DECaLS $r_{mag}$ of galaxies per redshift bin is shown as a square, along with the 16th and 84th percentiles. {\it Bottom panel:} The detection rate of \mgii absorbers host galaxies with equivalent width (\ew) for primary sample with $z < 1.6$ (square). The cumulative fraction for all detections, including primary and secondary samples, is given by the right-hand-side ordinates.}
    \label{fig:detection}
\end{figure}

\section{Results and Discussions} 
In this section, among \sample\ legitimate \mgii absorber galaxies from the present study, we use \detection\ galaxies with robust SED-based estimation of stellar mass and SFR to explore the nature of galaxies hosting \mgii absorbers. The examples of  \mgii absorber galaxies from our detection set are shown in  Figure~\ref{fig:collage}.
To strengthen the statistical analysis, we complement the above sample with \bmgiidetection\ galaxies from Paper II, with well-constrained SFR and stellar mass parameters. This resulted in a large data set of \primarydetection\ galaxies providing access to the largely unexplored impact parameter ranges of \primaryipmin\ to \primaryipmax\ kpc over a wide redshift range of \primaryzabsmin\ to \primaryzabsmax.

\begin{figure}
    \centering
    \includegraphics[width=0.49\textwidth]{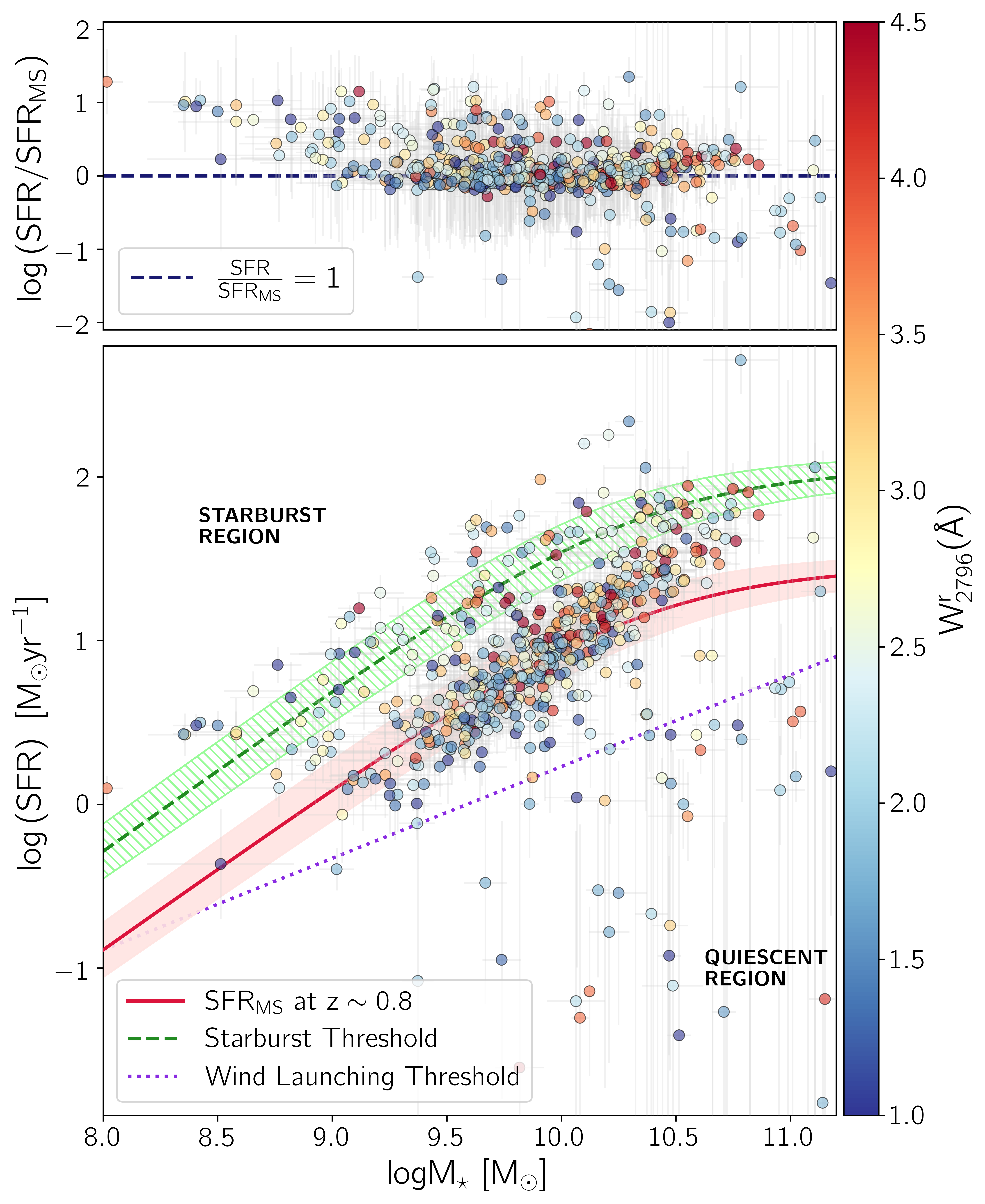}
    \caption{Star formation main sequence relation for \mgii absorber host galaxies. The symbols are colour-coded with the equivalent width. The shaded and hatched regions indicate the $1\sigma$ confidence interval. Additionally, the dotted line represents the SFR threshold for wind launching from \citet{Murray2011ApJ...735...66M}. {\it Top panel:} SFR of \mgii galaxies normalized by the main sequence star-forming galaxy at the respective redshift.}
    \label{fig:SFR_SM}
\end{figure}

\subsection{Detection rate of emission selected \mgii absorber galaxies}
Firstly, we estimate the detection rate of \mgii galaxies at close impact parameters of $\lesssim 20$kpc. The histogram in the top panel of Figure~\ref{fig:detection}, shows the detection fraction as a function of redshift for our primary sample (hatched), including only galaxies with good SED fit, and secondary sample (dashed) with \oii\ based selection. For $z \lesssim 1$, the average detection rate for the primary set is found to be 0.45\%, which reduces to $\lesssim 0.1$\% at $z > 1$. This is similar to the detection rate of 0.5\% from the SDSS-based selection from Paper II. The detection fraction increases by 22\% when we consider the secondary detection set.  In the bottom panel of Figure~\ref{fig:detection}, we show the detection fraction with rest-frame \ew. The detection rate increases from $\sim0.01$\% for \ew $\lesssim1$\AA\ to $\sim0.7$\% for ultra-strong \mgii absorbers with \ew $\ge 3$\AA\ for the primary set. However, the detection fraction mentioned above represents a lower limit, constrained by the finite fibre size, and limited imaging and spectroscopy sensitivity in identifying faint galaxies. In addition, the \ew\ is found to be $\ge 1$\AA\ for $\sim$\detectionewgeonerate\% of detections, which can be understood with the observed anti-correlation between \ew\ and impact parameter \citep{Chen2010ApJ...714.1521C,Nielsen2013ApJ...776..115N,Huang2021MNRAS.502.4743H}. It further suggests that the stronger absorbers likely trace the complex gas flows within the halos of galaxies, such as those driven by outflows or associated with accretion processes. 

\begin{figure}
    \centering
    \includegraphics[width=0.49\textwidth]{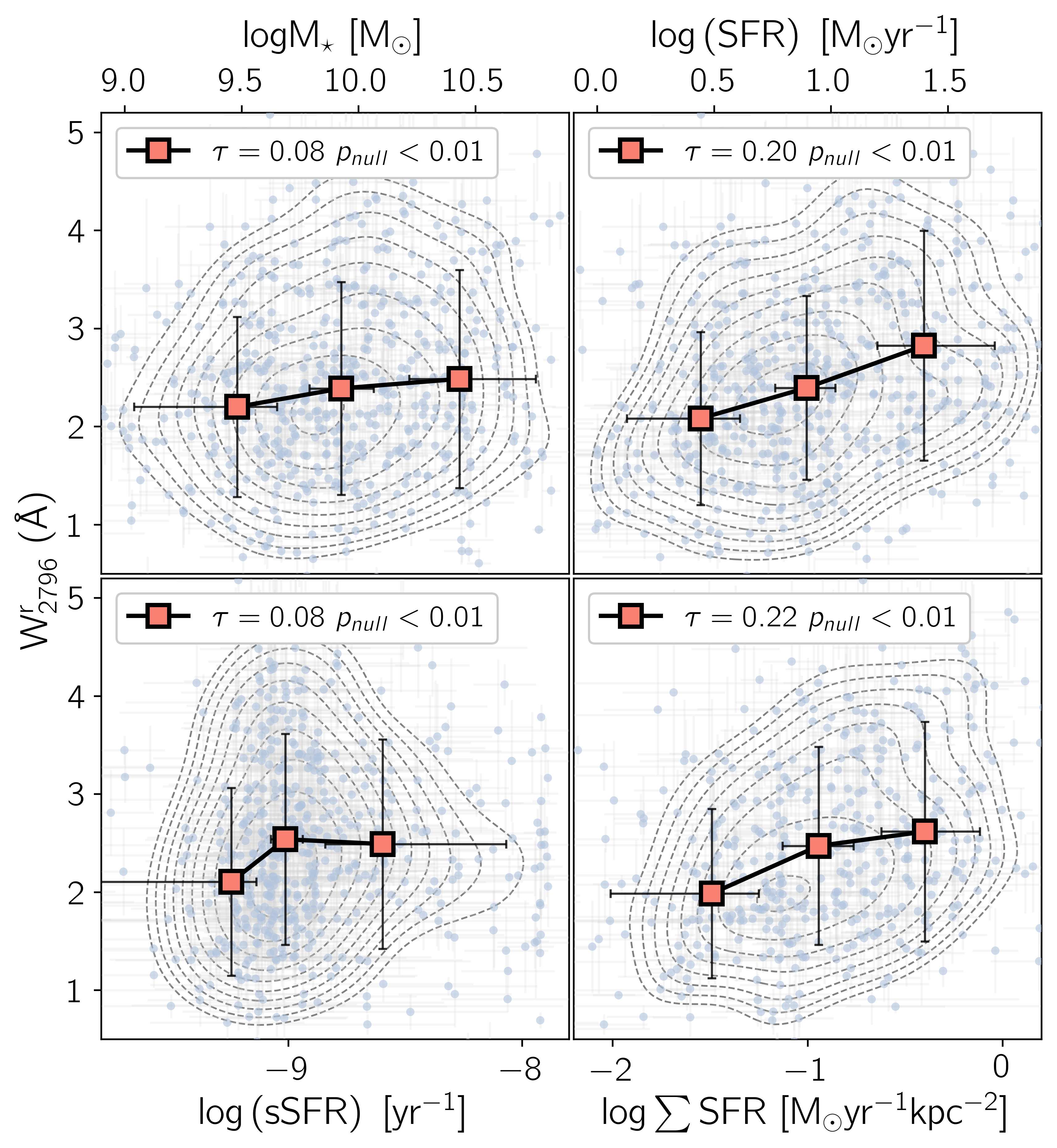}
    \caption{Distribution of \mgii equivalent width with the galaxy stellar mass (top-left), SFR (top-right), specific SFR (bottom-left), and star formation rate surface density (bottom-right). The square symbols represent the median values over three equal bins.}
    \label{fig:corr}
\end{figure}

\subsection{Physical properties of \mgii absorber host galaxies}
Using the largest set of spectroscopically identified \mgii absorber host galaxies at quasar proximity of $\lesssim $\primaryipmax\ kpc, we performed the multi-band SED modelling to derive the key physical parameters, such as stellar mass and SFR (see Section~\ref{sec:nebularemi}). We find that the \mgii galaxies, traced by strong nebular emission lines, probe a wide range of stellar mass ($\rm \primarymstarmin \le log (M_{\star}/  M_{\odot}) \le \primarymstarmax$), with an average $\rm \langle log\ M_{\star} \rangle =$  \primarymstarmedian\ $\rm M_{\odot}$. These galaxies form stars with $\rm \primarylogsfrmin \le \log SFR\ [M_{\odot} yr^{-1}] \le \primarylogsfrmax$, at a median  $\rm \langle SFR \rangle$ of \primarylogsfrmedian\ $\rm M_{\odot}\ yr ^{-1}$.  At first, we explore the \mgii absorber galaxies on the star formation main sequence plane (see Figure~\ref{fig:SFR_SM}). The stellar mass versus star formation distribution shows that the majority of \mgii galaxies are consistent with the main sequence relation for $\langle z \rangle \sim  0.8$  star-forming galaxies. The best fit relation from \citet{Popesso2023MNRAS.519.1526P} is shown as {\it red-solid} curve. The large fraction of galaxies with masses log$\rm (M_{\star}/M_{\odot}) > 10.5$ are at comparatively higher $z > 0.8$, leading to the small deviation from the main sequence relation. In the top panel of Figure~\ref{fig:SFR_SM}, we compare the SFR, normalized at the respective stellar mass and redshift based on the main sequence relation. Overall, $\sim$\primarystarburstrate\% of galaxies hosting  \mgii absorbers show a starburst nature, with an SFR of factor three to four higher than the main sequence galaxies \citep{Elbaz2018A&A...616A.110E}. For absorption-selected galaxies at the average $z \sim 0.8$, the starburst fraction is comparatively higher than 1-3\% of the general galaxy population \citep{Arango-Toro2025A&A...696A.159A}. We note that the large fraction of star-forming galaxies in our sample primarily reflects a sample selection bias, as the galaxies are selected based on the presence of strong \oii\ nebular emission.

Assuming that host galaxies do not harbor active galactic nuclei and stars as the ionizing source, we derived the \oii\ based SFR from the scaling relation given in Equation 5 of \citet[][]{Zhuang2019ApJ...882...89Z}.  The \oii\ based SFRs are found in the range of \primarysfroiimin\ to \primarysfroiimax\ $\rm M_{\odot}\ yr^{-1}$. Further,  considering the characteristic luminosity of $\rm log(L_{\star} [erg\ s^{-1}]$) = \logcharacteristicoiilum, for \oii\ luminosity function at average redshift  of $\langle z \rangle \sim \primaryzabsmedian$ in our sample, the \oii\ nebular line luminosity ranges between \primarylumoiinormmin–\primarylumoiinormmax\ $L^{\star}$\oii. It is worth recalling that the fiber aperture encompasses only a fraction of the total galaxy extent, resulting in the flux losses in the finite fiber size  \citep{Lopez2012MNRAS.419.3553L,Joshi2017MNRAS.471.1910J}. Therefore, the measured \oii\ emission line fluxes represent the lower limits. In addition, the derived SFRs are systematically smaller than the SED-based SFRs due to the fact that dust reddening is not taken into account. Henceforth, we used the SED-based SFR to study the characteristics of absorber host galaxies.  \par

In Figure ~\ref{fig:corr}, we explored the dependence of \mgii absorber strength on galaxies' key physical parameters, such as M$_{\star}$, SFR, specific star formation (sSFR), and star formation rate surface density ($\Sigma_{\rm  SFR} [\rm M_{\odot} yr^{-1} kpc^{-2}]$). The median \ew\ over three equal bins of M$_{\star}$, SFR, sSFR, and $\Sigma$SFR are shown as a square. The strength of \ew\ is mildly correlated with the stellar mass, with Kendall's $\tau$ correlation coefficient of $\tau_k$= 0.08 and a null probability of  $p_{null} \lesssim 0.01$. Analyzing the average \mgii absorption strength in the halo of $\sim$0.4 million star-forming galaxies (ELGs), \cite{Chen2025ApJ...981...81C} found that \ew\ strongly depends on stellar mass with  \ew\ $ \propto M_{\star}^{0.5}$ over $\rho <$100 kpc for galaxies with $\rm log (M_{\star}/M_{\odot}) \le 10$, and remain independent for higher mass galaxies \citep[see also,][]{Dutta2020MNRAS.499.5022D}. In contrast to the lack of correlation reported in Paper II, the statistically robust sample reveals a weak correlation. It can be partially explained by the higher covering fraction observed in the halos of high-$z$ galaxies \citep{Zahedy2019MNRAS.484.2257Z, Dutta2020MNRAS.499.5022D, Anand2021MNRAS.504...65A} and simulations \citep{Ho2020ApJ...904...76H, Nelson2020MNRAS.498.2391N, Ramesh2024MNRAS.528.3320R}. In addition, the large spread in \ew\ suggests that the CGM is not smooth, homogenous, and isotropic near the disk-halo interface, where strong outflows and gas accretion are evident \citep{Nielsen2013ApJ...776..115N,Tumlinson2017ARA&A..55..389T, Ponti2023A&A...670A..99P, Bisht2025MNRAS.tmp.1273B}.

In addition, the \ew\ is found to be more strongly correlated with the SFR with Kendall's $\tau$ correlation coefficient of $\tau_k$= 0.20 with a null probability of  $p_{null} \lesssim 0.01$ (see Figure~\ref{fig:corr})  .  The increase in the incidence of outflows with SFR suggests that the \mgii absorbers in our sample preferentially originate in galactic outflows \citep[see also,][]{Cicone2016A&A...588A..41C}. To further examine this relationship, we analyze \ew\ dependence on the $\Sigma_{\rm  SFR}$ in the {\it left-panel} of Figure~\ref{fig:corr}.  The average $\Sigma_{\rm SFR}$ is estimated by normalizing the SED-based SFR with the galaxy area measured from the best-fit ellipse parameters \citep{Dey2019AJ....157..168D}.  The \ew\ show a strong dependence on $\Sigma_{\rm  SFR}$ with Kendall's $\tau$ correlation coefficient of $\tau_k$= 0.22 and a null probability of $p_{null} \lesssim 0.01$. The majority \primaryhighsfrdensity/\primarydetection\ (\primaryhighsfrdensityrate\%) of galaxies have $\Sigma_{\rm SFR} \gtrsim 0.05 \rm M_{\odot} yr^{-1} kpc^{-2}$, sufficient to drive outflows of $10^4$K gas in galaxy halos out to 50–100 kpc \citep{Murray2011ApJ...735...66M, Hopkins2012MNRAS.421.3522H, Schroetter2024A&A...687A..39S, Chu2025MNRAS.536.1799R}. The higher star-formation rate of \mgii selected galaxies, along with a strong correlation with key physical parameters, i.e., SFR, $\Sigma$SFR, and M$_{\star}$, suggests the wind origin of \mgii absorbers \citep[see also,][]{Rubin2018ApJ...853...95R, Bordoloi2011ApJ...743...10B,Schroetter2024A&A...687A..39S}.

\subsection{Redshift evolution of cool gaseous halo}  
\label{sec:zevol}
The galaxies evolve significantly in morphology, having more compact and asymmetric profiles, and excess star-formation at high-$z$, than their low-$z$ counterparts of similar masses \citep{Behroozi2013ApJ...770...57B}.  In particular, galaxies at $z  \sim 2$ are seen actively forming stars and hosting super-galactic winds that are likely effective in replenishing the CGM with metal-enriched gas \citep{Nielsen2020ApJ...904..164N}.  Cosmological hydrodynamical simulations show that along with the galaxy, the gaseous halos undergo considerable changes in physical and dynamical states, such as density, temperature,  metallicity, velocity, and angular momentum. For instance,  high-$z$ gaseous haloes have nearly as much cool ($T < 10^5$ K) gas as hot ($T \ge 10^5$ K) gas out to $R_{200}$, while low-z haloes have five times more hot gas than cool gas \citep{Huscher2021MNRAS.500.1476H}.

\begin{figure}
    \centering
    \includegraphics[width=0.49\textwidth]{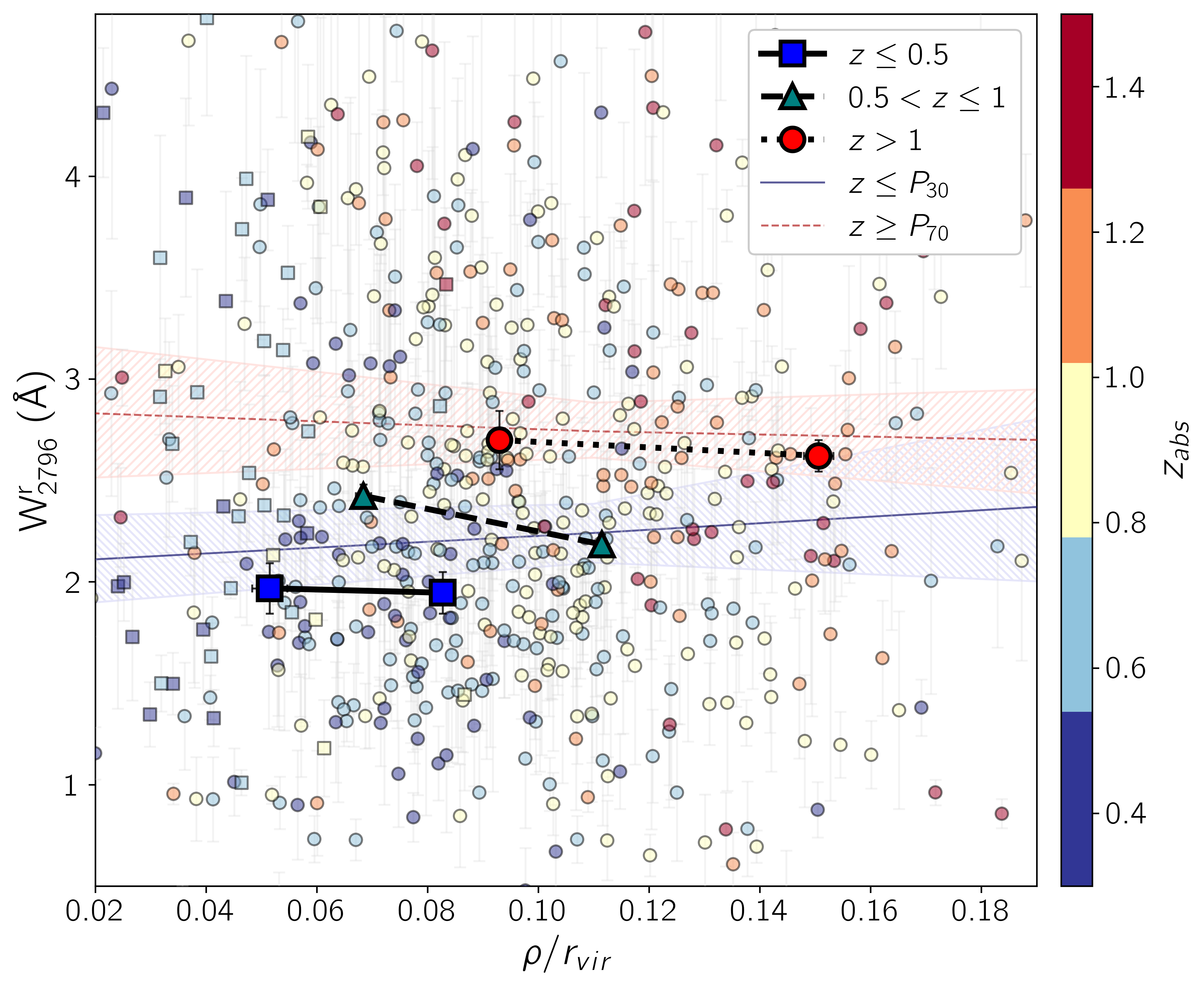}
    \caption{ Radial dependence of \mgii absorption strength. \mgii absorber rest-frame equivalent width versus impact parameter normalized by the galaxy's virial radius ($\rho/R_{vir}$). The best fit for high-$z$ ($z \sim \primarylepthirtyzabsmean$) and low-$z$ ($z \sim \primarygepseventyzabsmean$) subsets comprising the top and bottom 30\% of the sample is shown as {\it dotted} and {\it solid} line, with  $1\sigma$ uncertainty as hatched region. The average \ew for the $z \le 0.5$ ({\it blue-square}), $0.5 \le  z \le 1.0$ ({\it green-triangle}), and $z \ge 1$ ({\it red-circle}) subsets, for $\rho/R_{vir}$ bins shows  a clear evolution of mean absorption strength.}
    \label{fig:EW_D}
\end{figure}

From the observational perspective, \citet{Chen2012MNRAS.427.1238C} compared the average equivalent width of H{~\sc ii}, C{~\sc iv}, and \mgii absorption profiles from various absorption-line surveys conducted in the vicinities of galaxies over redshifts of $z \sim 0$  to 2.2.  They observed a good agreement in the spatial extent and mean absorption equivalent width between the CGM at low and high redshifts across the 11 Gyr of galaxy evolution. It primarily reflects a similar kinematics and volume filling factor of gaseous clumps in galactic haloes \citep[see also][]{Rudie2019ApJ...885...61R, Dutta2020MNRAS.499.5022D}. Using DESI-Y1 data of 2.5 million emission line galaxies (ELGs) and 1.4 million background quasars, \citet{Wu2025ApJ...983..186W} studied the evolution of cool gas, traced by \mgii absorption, in galaxy halos over $z =$ 0.75 to 1.65.  Conversely, they found that  ELGs at higher redshifts, $1.0 < z < 1.65$, harbor systematically higher gas fractions over spatial scales ranging from 20 kpc to 1 Mpc. Interestingly, in a follow-up analysis with the DESI-EDR data, \citet{Chen2025ApJ...981...81C} found no such trend over $\le 1$ Mpc scales. The excess absorber strength of high-$z$ galaxies in the stacking analysis  \citep{Wu2025ApJ...983..186W} is attributed to high-$z$ galaxies having higher $M_{\star}$ than the low-$z$ counterparts \citep[see also][]{Chen2025ApJ...981...81C}.  \par

\begin{figure*}
    \centering
    \includegraphics[width=0.95\textwidth]{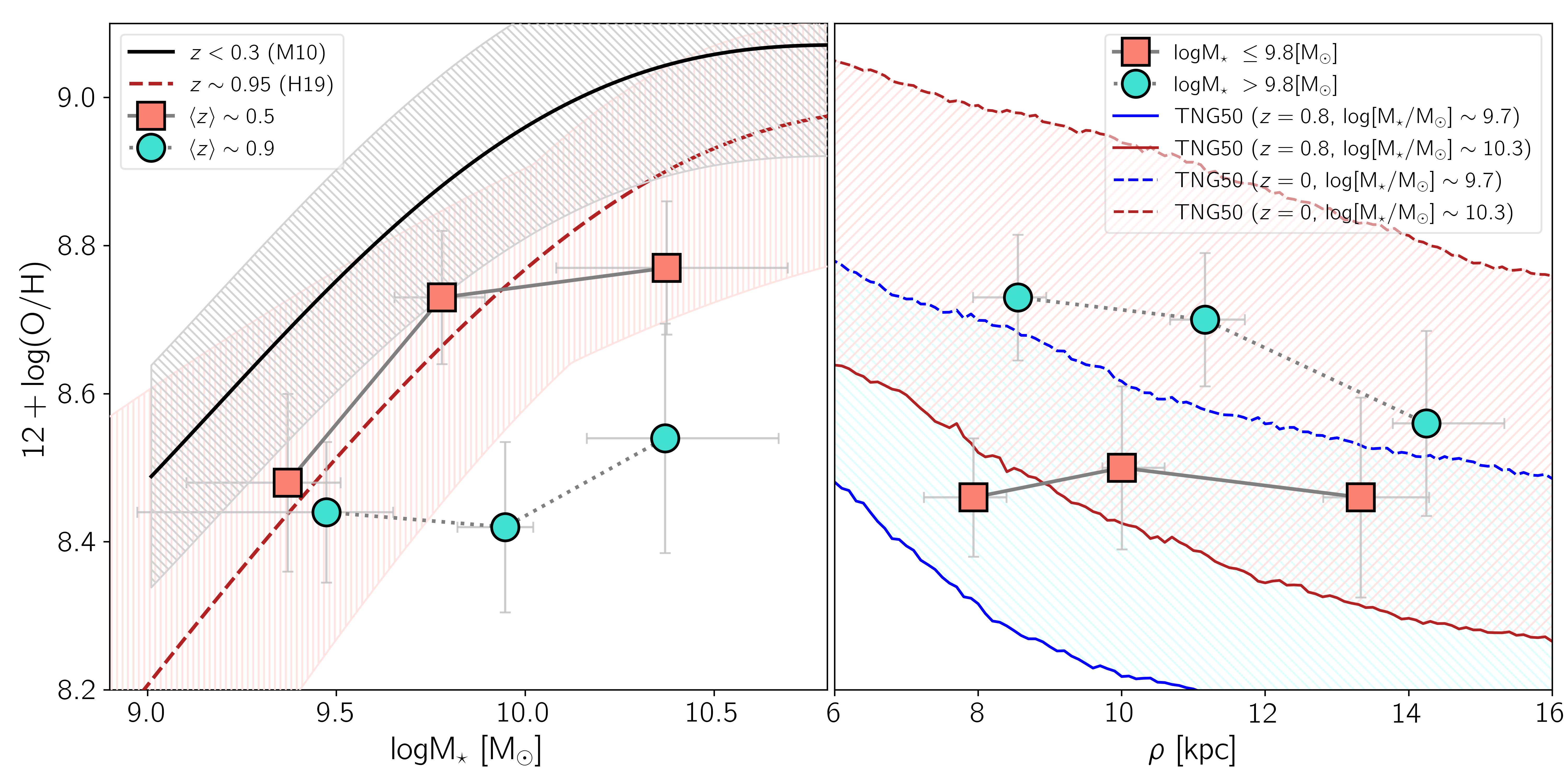}
    \caption{{\it Left panel:} Mass-metallicity relation for \mgii absorber host galaxies considering the top and bottom 30\% sample of redshift distribution with $\langle z \rangle$ of $\sim$0.51 (circle) and $\sim$0.86 (square), respectively.  
    The mass metallicty relation, median (solid and dashed curve) and 1-$\sigma$ scatter (hatched region at -45 and 90 degree), observed for $z \sim 0.1$ \citep{MannucciF.2010MNRAS.408.2115M} and $z \sim 0.95$ galaxies \citep{Huang2019ApJ...886...31H}, respectively. {\it Right panel:} The metallicty gradient with galactrocentric distance for the subset of low (square) and high (circle) mass galaxies with $\rm \langle log(M_{\star}/M_{\odot}) \rangle$ of 9.7, and 10.2,  respectively. The dashed and solid curves represent the gas-phase metallicity profiles for star-forming galaxies at $z \sim 0$ (red curve) and $z \sim 0.8$ (blue curve) in IllustrisTNG-50 \citep{Garcia2023MNRAS.519.4716G}.}
    \label{fig:r23}
\end{figure*}

In order to discern the evolution of cool gas in halos with redshift at small impact parameters of $< 20$kpc, in 
Figure \ref{fig:EW_D}, we show the \ew\ versus galactocentric distances from the galaxy, normalized by the virial radius ($\rho/R_{vir}$), for our entire sample, color coded with redshift. Further, we estimated the halo mass by converting the stellar mass into the halo mass using the stellar-to-halo mass relation from  \citet{Girelli2020A&A...634A.135G}. It is apparent that the absorbers at higher redshift show higher \ew\ than the low-$z$ counterparts.  We subdivide the sample into three redshift intervals of $z \le 0.5$, $0.5 \le z \le 1.0$, and $z > 1$,  consisting of \primarylowzabs, \primarymidzabs, and \primaryhighzabs\ galaxies, respectively. Despite higher-$z$ absorbers probing relatively larger galactocentric distances, due to larger projected fiber area, a systematic increase in average \ew\ of \primarylowzabsewmean\AA, \primarymidzabsewmean\AA, and \primaryhighzabsewmean\AA\ (each with mean error of $0.1$\AA) is observed with $z$. Further, we compared the average \ew\ as a function of $\rho/R_{vir}$ over the above three $z$ bins. The absorbers at $z > 1$ clearly show higher \ew\ ({\it red-circle}) compare to absorbers at $z \le 0.5$ ({\it blue-square}), $0.5 \le z \le 1.0$ ({\it green-triangle}). 

Furthermore, we model the \mgii absorption strength as a function of $\rho$ by fitting a log-linear model  log\ew\ (\AA) = $\alpha$ + $\beta \times \rho$ (kpc), with a likelihood function given in equation 7 of \citet[][]{Chen2010ApJ...714.1521C}. For this,  we generate the high-$z$ and low-$z$ subsets using the top and bottom 30\% of the sample, each consisting of \primarylepthirtyzabs\ absorbers with average $z$ of \primarylepthirtyzabsmean, and \primarygepseventyzabsmean, respectively. The best-fit parameter for the high-$z$ and low-$z$ subsets are $\alpha = \primarygepszabsalpha^{+\primarygepszabsalphaerrhigh}_{-\primarygepszabsalphaerrlow}, \ \beta = \primarygepszabsbeta^{+\primarygepszabsbetaerrhigh}_{-\primarygepszabsbetaerrlow}$ and $\alpha = \primaryleptzabsalpha^{+\primaryleptzabsalphaerrhigh}_{-\primaryleptzabsalphaerrlow}, \ \beta = \primaryleptzabsbeta^{+\primaryleptzabsbetaerrhigh}_{-\primaryleptzabsbetaerrlow}$, respectively. Figure~\ref{fig:EW_D} shows the best-fit log-linear model, along with a 1$\sigma$ uncertainty in the shaded region. The enhanced mean absorber strength is evident in high-$z$ halos with a mild anti-correlation with $\rho/R_{vir}$. It reiterates the higher gas covering fraction and non-homogeneous nature of CGM due to strong outflows and strong accretion events near the disk, which are more prevalent at higher redshift, where galaxies exhibit higher SFR and sSFR (\citep{Tumlinson2017ARA&A..55..389T,Dutta2020MNRAS.499.5022D,Ponti2023A&A...670A..99P,  Ramesh2024MNRAS.528.3320R,  Bisht2025MNRAS.tmp.1273B,Das2025A&A...695A.207D, Joshi2025A&A...695A.206J}.  \par

We note that the high-$z$ and low-$z$ subsets differ in stellar masses. To address the contribution of stellar mass to the observed excess \ew\ at high-$z$ halos \citep{Wu2025ApJ...983..186W, Chen2025ApJ...981...81C},  we constructed two subsets by selecting the galaxies within the range of $\rm log (M_{\star}/  M_{\odot}) > 9.9$, comprising \gepszabshm\ and \leptzabshm\ galaxies in high-$z$ and low-$z$ bins, respectively. A Kolmogorov-Smirnov (KS) test statistic of $D=\lzhzmassstat$ at $p_{null}=\lzhzmassstatpvalue$, indicates no significant difference in $M_{\star}$ distribution of the two subsets. The average \ew\ for the mass-matched high-$z$ and low-$z$ subsets are $\hzhmewmean \pm \hzhmewstd$\AA, and $\lzhmewmean \pm \lzhmewstd$ \AA, respectively. The KS-test null probability of the sample being drawn from the parent \ew\ distribution is rejected with $D=\lzhzmgiiewstat$ at $p_{null}=\lzhzmgiiewstatpvalue$. This suggests that the stronger absorption observed in the halos of high-z galaxies reflects an evolution of cool CGM gas with redshift, particularly at small impact parameters. The median SFR of \hzhmsfrmedian\ and \lzhmsfrmedian\ $\rm M_{\odot}\ yr^{-1}$ for the high and low $z$ subsets is comparable and a factor of half of the main sequence galaxies, respectively, underscores the role of feedback processes in enriching the CGM.

\par

\begin{figure*}
    \centering
    \includegraphics[width=0.95\textwidth]{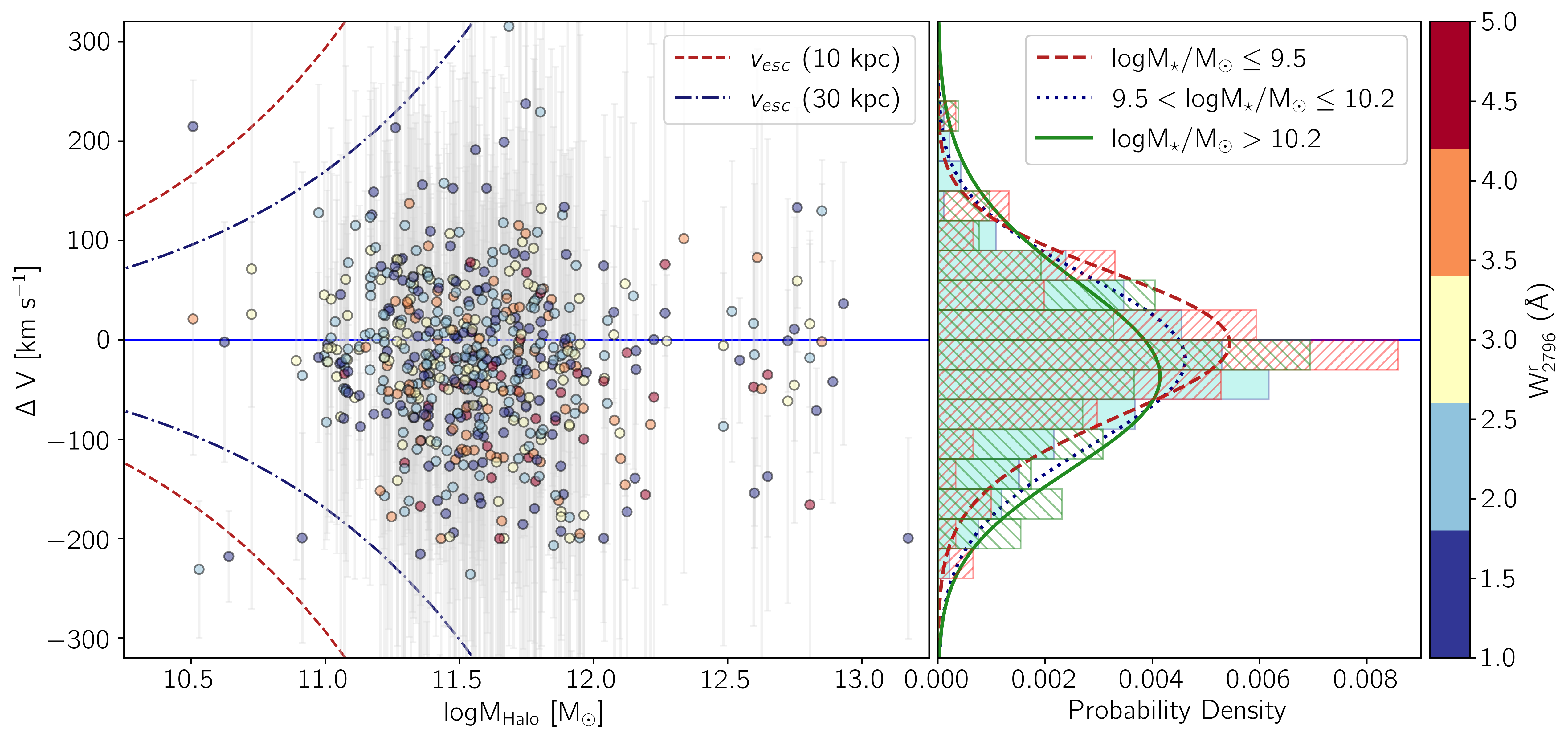}
    \caption{
    {\it Left Panel:} Relative \mgii absorption velocities to the galaxy systematic redshifts as a function of inferred dark matter halo mass; color coded with \ew. The line-of-sight projected halo escape velocities at distances 10 and 30 kpc, assuming an NFW profile of dark matter halos, are marked as  {\it dashed} and {\it dot-dashed} curves.  {\it Right Panel:}  The area normalized histogram of relative velocity of \mgii absorbers in star-forming galaxies for three mass bins of $\rm log(M_{\star}/M_{\odot} \le 9.5$ (hatched at 45 degree),   $\rm 9.5 \le  log(M_{\star}/M_{\odot}) \le 10.2$ (hatched at -45 degree), and $\rm log(M_{\star}/M_{\odot} \ge 10.2$ (filled histogram). The Gaussian profile for the high mass bin (solid curve) shows higher velocity dispersion with $\sigma =\primaryhmdvsigma$ \kms\ than $\sigma =\primarylmdvsigma$ \kms, $\sigma =\primarymmdvsigma$ \kms\  for the respective low mass bins of  $\rm log(M_{\star}/M_{\odot}) \le 9.5$ ({\it dashed curve}), and   $\rm 9.5 \le log(M_{\star}/M_{\odot}) \le 10.2$ ({\it dotted curve}).}
    \label{fig:DELTAV_MHALO}
\end{figure*}

\subsection{Gas-phase metallicity} 
The local galaxies exhibit a tight correlation, with a scatter of only $\sim$0.1 dex,  of increasing gas-phase metallicity with increasing galaxy stellar mass across several orders of magnitude, suggesting a close link between metal content and galaxy evolution \citep{Tremonti2004ApJ...613..898T, Finkelstein2012ApJ...756..164F}. The mass-metallicity relation is observed out to $z \sim$ 4  with a decrease in the overall metallicity of galaxies with increasing redshift \citep{Zahid2011ApJ...730..137Z,Zahid2014ApJ...791..130Z}. The metal-enriched outflows and gas accretion play a crucial role in shaping the observed fundamental metallicity relation with massive and high star-forming galaxies possessing higher metallicity than their lower mass and star-formation counterparts \citep{MannucciF.2010MNRAS.408.2115M, LequeuxJ.1979A&A....80..155L,Tremonti2004ApJ...613..898T}. In this section, we explore whether the fundamental relation holds for \mgii absorber host galaxies that are selected based on \oii\ nebular emission line.

At first, we explore the redshift evolution of the gas-phase metallicity of absorption-selected galaxies. For this, we restrict our sample to galaxies with $z \leq 0.93$, ensuring that the \oiiiab\ emission line falls within the DESI spectral range. To assess the evolution with redshift,  we divide the sample into two redshift bins while considering the absorbers from the top and bottom 30\% of redshift distributions, with  $\langle z \rangle$ of \rttlzzabsmedian\ and \rtthzzabsmedian, respectively. We generate the spectral stack for equally divided three mass bins for both low-z and high-z subsets, and estimate the \oiiab, \oiiiab, \hbeta line fluxes and metallicity-sensitive emission-line ratio R23 \citep{Alloin1979A&A....78..200A}. The gas-phase oxygen abundance $\rm 12 + log(O/H)$ is derived by employing a non-parametric approach based on \citet{Langeroodi2024arXiv240907455L}, utilizing the genesis-metallicity tool\footnote{\href{https://github.com/langeroodi/genesis_metallicity}{https://github.com/langeroodi/genesis\_metallicity}}. It captures the real empirical trend based on a calibration sample of 1510 galaxies at $0 < z < 10$ with direct-method metallicity measurements and kernel density estimate in the 4-dimensional space of O2 = \oiiab/\hbeta;  O3 = \oiiib/\hbeta; \hbeta equivalent width, and gas-phase metallicity. The left panel of Figure~\ref{fig:r23} shows the distribution of metallicity as a function of stellar mass. It is evident, alike the general galaxy population, the metallicity for \mgii absorber host galaxies evolves with stellar mass. We note that at a median absorber redshift of $\langle z \rangle \sim 0.8$, the estimated metallicity for the \mgii absorbers shows a small departure from the mass-metallicity relation for the general galaxy population \citep{Huang2019ApJ...886...31H}, that most likely results from the flux loss in the fiber, which only covers a portion of the galaxy. Furthermore, the galaxies at higher $\langle z \rangle \sim 0.8$ show lower metallicity ({\it circle}) than the lower $\langle z \rangle \sim 0.5$ systems ({\it square}). As prior studies largely lack evidence for the redshift evolution of \ew\ with galactocentric distance, it likely imply that CGM activity predominantly takes place in the inner regions on shorter timescales, where later $\rho/R_{\rm vir}$ studies are less sensitive.

The DESI fibers, centered on the quasars, sample a broad range of impact parameters (i.e., \primaryipmin–\primaryipmax\ kpc). Given the finite fiber size, it partially covers the absorber galaxy. In this configuration, for galaxies at larger impact parameters, the fiber primarily probes the galaxy's halo or the disk–halo interface, whereas for smaller impact parameters, it captures the galaxy disk. In Figure~\ref{fig:r23}, we present the gas-phase metallicity as a function of galactocentric radius for two stellar mass bins, with $\rm log( M_{\star}/M_{\odot})$ ranging from \rttlmmstarmin–\rttlmmstarmax\ and \rtthmmstarmin–\rtthmmstarmax, and corresponding  $\rm \langle log ( M_{\star} /M_{\odot})\rangle$ of \rttlmmstarmedian\ and \rtthmmstarmedian, respectively. The higher-mass galaxies exhibit a declining metallicity gradient with increasing radius, while the lower-mass galaxies show an approximately flat metallicity profile. We further compare it with radial gas-phase metallicity profiles for star-forming galaxy population at redshifts of $z \sim 0$ and  $z \sim 0.8$ from the TNG50 \citep{Garcia2023MNRAS.519.4716G}.  It is evident that the galaxies in the TNG50 simulation exhibit a decline in metallicity with increasing radius. In addition, the metallicity estimates for \mgii absorber host galaxies for high mass bins are consistent with the star-forming galaxy in TNG50. Conversely, no such radial gradient is observed for the low mass galaxies. \citet{Decataldo2024A&A...685A...8D} analysed the origin of the metal-enriched cold CGM in the smoothed particle hydrodynamics cosmological zoom-in simulation Eris2k. They show that the outflows have a major contribution to the cold CGM gas budget at $z <1$, with almost 50\% of the hot gas cooling in outflow. The higher gas-phase metallicty in galaxy halos and higher star-formation rates reaffirm that galaxies exhibiting strong \mgii absorption are typically star-forming systems characterized by ubiquitous outflows.

\subsection{Absorber kinematics}
The strong \mgii absorbers in low-resolution spectra largely trace the multiple gas cloud clumps \citep{Churchill2013ApJ...763L..42C}. To map the underlying motion, we estimate the line-of-sight velocity offset ($\Delta V$) between the systemic galaxy redshift based on \oii\ emission and \mgii absorber cloud. Figure~\ref {fig:DELTAV_MHALO} shows $\Delta V$ versus galaxy halo mass (see Section~\ref{sec:zevol}). 
% The halo masses for each absorber redshifts were derived from the stellar-to-halo mass relation given by \citet{Girelli2020A&A...634A.135G}, assuming a spherically symmetric Navarro–Frenk–White dark matter halo profile \citep{Navarro1996ApJ...462..563N}. 
The $\Delta V$ ranges between \primarydvmin\ to \primarydvmax ~\kms with a dispersion of \primarydvstd\ $\rm km\ s^{-1}$ and  mean velocity difference of  \primarydvmean\ $\rm km\ s^{-1}$. The escape velocities at an average impact parameter of 10 kpc and 30 kpc with respect to mass are shown as {\it dashed } and {\it dot-dashed } curves, respectively. It is evident that the majority of the \mgii absorber clouds have velocities lower than the expected projected escape velocities, indicating that they are bound to dark matter potential and are likely recycled back into the galaxy as galactic fountains. It is consistent with high-resolution CGM-ZOOM simulations showing halo gas undergoing multiple recycling through galactic fountains \citep{Suresh2019MNRAS.483.4040S}.

Using galaxy-quasar pairs of 211 isolated and 43 non-isolated galaxies at  $\langle z \rangle \sim 0.21$ and projected distances of $\langle \rho \rangle$ 86 kpc, \citet{Huang2021MNRAS.502.4743H} found that the velocity distribution of absorbing gas depends on host galaxy properties. For star-forming galaxies, a 20\% excess of velocity dispersion ($\sigma$) is observed for high mass ($\rm \langle log(M_{\star}/M_{\odot}) \rangle \sim$ 10.7) galaxies than low mass (\rm  $\rm \langle  log(M_{\star}/M_{\odot}) \rangle \sim$ 9.7) subset with  $\sigma \sim $ 83 \kms\ and 67\kms, respectively. In the $right$ panel of Figure~\ref{fig:DELTAV_MHALO}, we show the velocity dispersion for three mass bins of $\rm log(M_{\star}/M_{\odot}) \le 9.5$,  $\rm 9.5 \le  log(M_{\star}/M_{\odot}) \le 10.2$, $\rm log(M_{\star}/M_{\odot}) > 10.2$, consisting of \primarydvlm, \primarydvmm, and \primarydvhm\ galaxies, respectively. For the above three subsets, the mean velocity offset is found to be $\Delta V \sim$ \primarylmdvmean, \primarymmdvmean, and \primaryhmdvmean\ \kms. Similarly, the velocity dispersion of \mgii absorbers positively correlates with the increasing stellar mass with $\sigma$ = \primarylmdvsigma, \primarymmdvsigma, and \primaryhmdvsigma\ \kms, respectively. We further estimate the 68th percentile confidence interval using bootstrap analysis. The $\sigma$ of $\primaryhmdvsigmamin-\primaryhmdvsigmamax$~\kms for high mass bins of $\rm log(M_{\star}/M_{\odot}) > 10.2$ is higher by $\sim30$\% than $\sigma$ = $\primarylmdvsigmamin-\primarylmdvsigmamax$\kms for low mass bins of $\rm log(M_{\star}/M_{\odot}) < 9.5$, suggest the positive correlation with the halo mass. The higher velocity dispersion in massive galaxies can be understood as a result of high star formation activity regulating the strong winds or gas accretion in the massive halos. In addition, gas associated with satellite galaxies, including gas ejected in winds or ram pressure stripping, can exist in halos for an extended period of time, and can also smoothly accrete by the central galaxy, leading to excess velocity dispersion \citep{Hafen2019MNRAS.488.1248H}. This scenario is further corroborated by the CGM in TNG50, where $\sim$40\% of the 
H{~\sc i} absorbers fraction with column density \rm logN(H{~\sc i}) [$cm^{-2}$] $ > 16$ and $\Delta V$ of 70-250~\kms\ is associated with the satellites around the central galaxy of $\rm M_{\star} < 10^9 M_{\odot}$ \citep{weng2024physical}.

\section{Conclusion}
We investigated the nature of \mgii absorbers galaxies over a redshift range of $\primaryzabsmin \le z \le \primaryzabsmax$, at largely unexplored galactrocentric distances of $\lesssim 20$kpc. The physical properties, including stellar mass, SFR, and $\Sigma$SFR for a large set of \detection\ absorber host galaxies identified in the present study, combined with \bmgiidetection\ galaxies from Paper II,  compared with the associated absorber properties, led to the following key results:

\begin{enumerate}
\item The average detection rate of \mgii absorber hosts based on strong nebular emission line in the DESI spectroscopic surveys is $\sim$\detectionrate\% below $z<1$. It is preferentially higher towards the strong absorbers with \detectionewgetwo/\detection\  (\detectionewgetworate\%) system having \ew $\ge 2$\AA. The detection rate increases by a factor of 30, from less than 0.02\% to 0.6\%, for low \ew ($\le 0.5$\AA) and ultra-strong absorbers. The \mgii absorber galaxies trace a wide range of stellar mass with  $\rm \primarymstarmin \le log(M_{\star}/M_{\odot}) \le \primarymstarmax$, and star-formation rates of $\rm \primarylogsfrmin \le log SFR \le \primarylogsfrmax$. These galaxies follow the star formation main sequence, with \primarystarburstrate\% of systems exhibiting a starburst nature.

\item  The \ew\ remains near constant as a function of galactrocentric distances, indicating a patchy structure of CGM and high covering fraction of cool gas in galaxy halos extending out to impact parameters of $\lesssim 20$~kpc.  The mean absorber strength rises from \primarylowzabsewmean\AA\ to \primaryhighzabsewmean\AA\ for the galaxies at median $\langle z \rangle $  of $\sim$\primarylowzabsmean\ and $\sim$\primaryhighzabsmean, suggesting the cool gas in halos evolves with redshift. Furthermore, a positive correlation between the physical properties of galaxies ($\rm M_{\star}$, SFR, and $\Sigma \rm SFR$) and absorption strength (\ew) suggests that the distribution of the metal-enriched cool gas in the CGM significantly depends on the properties of the central galaxy. 

\item The gas phase metallicity of absorption-selected galaxies evolves with stellar mass and redshift, following the mass metallicity relation of general galaxies. For the massive galaxies with $\rm \langle log(M_{\star}/M_{odot}) \rangle \sim \rtthmmstarmedian$, the gas phase metallicity declines as a function of radius, whereas the low mass galaxies $\rm \langle logM_{\star} \rangle \sim \rttlmmstarmedian$ show a relatively shallow evolution. This trend is consistent with the  TNG50 galaxies at z$\sim 0.8$. It suggests that many \mgii absorber galaxies are representative of ordinary star-forming galaxies.

\item We find that the velocity dispersion of the cool gas increases with halo mass. The wide range of line of sight velocities (\primarydvmin\ to \primarydvmax ~\kms) with an average of \primarydvmean\ \kms\ between galaxy and absorbing gas, and dispersion of \primarydvstd~\kms, highlights the dynamical behavior of CGM,  largely contributed by the gas accretion and strong winds in massive halos with high star formation activity. These gas clouds are largely bound to the dark matter potential. 

\end{enumerate}

In this study, we have investigated the properties of \mgii absorber host galaxies, specifically focusing on systems detected based on strong \oii\ nebular emission and DECaLS imaging. However, it is important to note that the current sample, by design, is biased toward galaxies with high \oii\ luminosity, and does not represent the general galaxy population of \mgii absorbers. The follow-up study of a luminosity-unbiased sample from deep HSC Subaru imaging \citep[see also,][]{Joshi2025A&A...695A.206J}, will help to get insight into the diversity of mechanisms responsible for the origin of \mgii absorbers.

\section{Data availability} The \mgii absorber host galaxy catalogue for \detection\ primary and \detectionsec\ secondary detection set is only available in electronic form at the CDS via anonymous ftp to cdsarc.u-strasbg.fr (130.79.128.5) or via \href{http://cdsweb.u-strasbg.fr/cgi-bin/qcat?J/A+A/}{http://cdsweb.u-strasbg.fr/cgi-bin/qcat?J/A+A/}.

\begin{acknowledgements}
%We acknowledge the anonymous referee for the constructive comments.  
%
LCH was supported by the National Science Foundation of China (12233001) and the China Manned Space Program (CMS-CSST-2025-A09). 

We acknowledge the contribution of Alex M. Garcia from the Department of Astronomy at the University of Virginia for providing the Illustris TNG50 simulation data on R23 metallicity gradients.

Funding for the Sloan Digital Sky Survey IV has been provided by the Alfred P. Sloan Foundation, the U.S. Department of Energy Office of Science, and the Participating Institutions. SDSS-IV acknowledges support and resources from the Center for High Performance Computing at the University of Utah. The SDSS website is www.sdss4.org.

SDSS-IV is managed by the Astrophysical Research Consortium for the Participating Institutions of the SDSS Collaboration including the Brazilian Participation Group, the Carnegie Institution for Science, Carnegie Mellon University, Center for Astrophysics | Harvard \& Smithsonian, the Chilean Participation Group, the French Participation Group, Instituto de Astrof\'isica de Canarias, The Johns Hopkins University, Kavli Institute for the Physics and Mathematics of the Universe (IPMU) / University of Tokyo, the Korean Participation Group, Lawrence Berkeley National Laboratory, Leibniz Institut f\"ur Astrophysik Potsdam (AIP),  Max-Planck-Institut f\"ur Astronomie (MPIA Heidelberg), Max-Planck-Institut f\"ur Astrophysik (MPA Garching), Max-Planck-Institut f\"ur Extraterrestrische Physik (MPE), National Astronomical Observatories of China, New Mexico State University, New York University, University of Notre Dame, Observat\'ario Nacional / MCTI, The Ohio State University, Pennsylvania State University, Shanghai Astronomical Observatory, United Kingdom Participation Group, Universidad Nacional Aut\'onoma de M\'exico, University of Arizona, University of Colorado Boulder, University of Oxford, University of Portsmouth, University of Utah, University of Virginia, University of Washington, University of Wisconsin, Vanderbilt University, and Yale University.

The DESI Legacy Imaging Surveys consist of three individual and complementary projects: the Dark Energy Camera Legacy Survey (DECaLS), the Beijing-Arizona Sky Survey (BASS), and the Mayall z-band Legacy Survey (MzLS). DECaLS, BASS and MzLS together include data obtained, respectively, at the Blanco telescope, Cerro Tololo Inter-American Observatory, NSF’s NOIRLab; the Bok telescope, Steward Observatory, University of Arizona; and the Mayall telescope, Kitt Peak National Observatory, NOIRLab. NOIRLab is operated by the Association of Universities for Research in Astronomy (AURA) under a cooperative agreement with the National Science Foundation. Pipeline processing and analyses of the data were supported by NOIRLab and the Lawrence Berkeley National Laboratory (LBNL). Legacy Surveys also uses data products from the Near-Earth Object Wide-field Infrared Survey Explorer (NEOWISE), a project of the Jet Propulsion Laboratory/California Institute of Technology, funded by the National Aeronautics and Space Administration. Legacy Surveys was supported by: the Director, Office of Science, Office of High Energy Physics of the U.S. Department of Energy; the National Energy Research Scientific Computing Center, a DOE Office of Science User Facility; the U.S. National Science Foundation, Division of Astronomical Sciences; the National Astronomical Observatories of China, the Chinese Academy of Sciences and the Chinese National Natural Science Foundation. LBNL is managed by the Regents of the University of California under contract to the U.S. Department of Energy. The complete acknowledgments can be found at https://www.legacysurvey.org/acknowledgment/.

\end{acknowledgements}

%
%-------------------------------------------------------------
%               Appendices have to be placed at the end, after
%                                        \end{thebibliography}
%-------------------------------------------------------------

% \begin{thebibliography}{}
% \end{thebibliography}

%
%

\bibliographystyle{aa} % style aa.bst
\bibliography{aa} % your references Yourfile.bib

\begin{thebibliography}{88}
\expandafter\ifx\csname natexlab\endcsname\relax\def\natexlab#1{#1}\fi

\bibitem[{{Afruni} {et~al.}(2021){Afruni}, {Fraternali}, \& {Pezzulli}}]{Afruni2021MNRAS.501.5575A}
{Afruni}, A., {Fraternali}, F., \& {Pezzulli}, G. 2021, \mnras, 501, 5575

\bibitem[{{Alloin} {et~al.}(1979){Alloin}, {Collin-Souffrin}, {Joly}, \& {Vigroux}}]{Alloin1979A&A....78..200A}
{Alloin}, D., {Collin-Souffrin}, S., {Joly}, M., \& {Vigroux}, L. 1979, \aap, 78, 200

\bibitem[{{Anand} {et~al.}(2021){Anand}, {Nelson}, \& {Kauffmann}}]{Anand2021MNRAS.504...65A}
{Anand}, A., {Nelson}, D., \& {Kauffmann}, G. 2021, \mnras, 504, 65

\bibitem[{{Arango-Toro} {et~al.}(2025){Arango-Toro}, {Ilbert}, {Ciesla}, {Shuntov}, {Aufort}, {Mercier}, {Laigle}, {Franco}, {Bethermin}, {Le Borgne}, {Dubois}, {McCracken}, {Paquereau}, {Huertas-Company}, {Kartaltepe}, {Casey}, {Akins}, {Allen}, {Andika}, {Brinch}, {Drakos}, {Faisst}, {Gozaliasl}, {Harish}, {Kaminsky}, {Koekemoer}, {Kokorev}, {Liu}, {Magdis}, {Martin}, {Moutard}, {Rhodes}, {Rich}, {Robertson}, {Sanders}, {Sheth}, {Talia}, {Toft}, {Tresse}, {Valentino}, {Vijayan}, \& {Weaver}}]{Arango-Toro2025A&A...696A.159A}
{Arango-Toro}, R.~C., {Ilbert}, O., {Ciesla}, L., {et~al.} 2025, \aap, 696, A159

\bibitem[{{Bacon} {et~al.}(2023){Bacon}, {Brinchmann}, {Conseil}, {Maseda}, {Nanayakkara}, {Wendt}, {Bacher}, {Mary}, {Weilbacher}, {Krajnovi{\'c}}, {Boogaard}, {Bouch{\'e}}, {Contini}, {Epinat}, {Feltre}, {Guo}, {Herenz}, {Kollatschny}, {Kusakabe}, {Leclercq}, {Michel-Dansac}, {Pello}, {Richard}, {Roth}, {Salvignol}, {Schaye}, {Steinmetz}, {Tresse}, {Urrutia}, {Verhamme}, {Vitte}, {Wisotzki}, \& {Zoutendijk}}]{Bacon2023A&A...670A...4B}
{Bacon}, R., {Brinchmann}, J., {Conseil}, S., {et~al.} 2023, \aap, 670, A4

\bibitem[{{Behroozi} {et~al.}(2013){Behroozi}, {Wechsler}, \& {Conroy}}]{Behroozi2013ApJ...770...57B}
{Behroozi}, P.~S., {Wechsler}, R.~H., \& {Conroy}, C. 2013, \apj, 770, 57

\bibitem[{{Bielby} {et~al.}(2017){Bielby}, {Crighton}, {Fumagalli}, {Morris}, {Stott}, {Tejos}, \& {Cantalupo}}]{Bielby2017MNRAS.468.1373B}
{Bielby}, R., {Crighton}, N.~H.~M., {Fumagalli}, M., {et~al.} 2017, \mnras, 468, 1373

\bibitem[{{Bisht} {et~al.}(2025){Bisht}, {Sharma}, {Dutta}, \& {Nath}}]{Bisht2025MNRAS.tmp.1273B}
{Bisht}, M.~S., {Sharma}, P., {Dutta}, A., \& {Nath}, B.~B. 2025, \mnras [\eprint[arXiv]{2411.17173}]

\bibitem[{{Bolton} {et~al.}(2004){Bolton}, {Burles}, {Schlegel}, {Eisenstein}, \& {Brinkmann}}]{Bolton2004AJ....127.1860B}
{Bolton}, A.~S., {Burles}, S., {Schlegel}, D.~J., {Eisenstein}, D.~J., \& {Brinkmann}, J. 2004, \aj, 127, 1860

\bibitem[{{Bordoloi} {et~al.}(2011){Bordoloi}, {Lilly}, {Knobel}, {Bolzonella}, {Kampczyk}, {Carollo}, {Iovino}, {Zucca}, {Contini}, {Kneib}, {Le Fevre}, {Mainieri}, {Renzini}, {Scodeggio}, {Zamorani}, {Balestra}, {Bardelli}, {Bongiorno}, {Caputi}, {Cucciati}, {de la Torre}, {de Ravel}, {Garilli}, {Kova{\v{c}}}, {Lamareille}, {Le Borgne}, {Le Brun}, {Maier}, {Mignoli}, {Pello}, {Peng}, {Perez Montero}, {Presotto}, {Scarlata}, {Silverman}, {Tanaka}, {Tasca}, {Tresse}, {Vergani}, {Barnes}, {Cappi}, {Cimatti}, {Coppa}, {Diener}, {Franzetti}, {Koekemoer}, {L{\'o}pez-Sanjuan}, {McCracken}, {Moresco}, {Nair}, {Oesch}, {Pozzetti}, \& {Welikala}}]{Bordoloi2011ApJ...743...10B}
{Bordoloi}, R., {Lilly}, S.~J., {Knobel}, C., {et~al.} 2011, \apj, 743, 10

\bibitem[{{Bouch{\'e}} {et~al.}(2012){Bouch{\'e}}, {Hohensee}, {Vargas}, {Kacprzak}, {Martin}, {Cooke}, \& {Churchill}}]{Bouche2012MNRAS.426..801B}
{Bouch{\'e}}, N., {Hohensee}, W., {Vargas}, R., {et~al.} 2012, \mnras, 426, 801

\bibitem[{{Bouch{\'e}} {et~al.}(2007){Bouch{\'e}}, {Murphy}, {P{\'e}roux}, {Davies}, {Eisenhauer}, {F{\"o}rster Schreiber}, \& {Tacconi}}]{Bouche2007ApJ...669L...5B}
{Bouch{\'e}}, N., {Murphy}, M.~T., {P{\'e}roux}, C., {et~al.} 2007, \apjl, 669, L5

\bibitem[{{Calzetti} {et~al.}(1994){Calzetti}, {Kinney}, \& {Storchi-Bergmann}}]{Calzetti1994ApJ...429..582C}
{Calzetti}, D., {Kinney}, A.~L., \& {Storchi-Bergmann}, T. 1994, \apj, 429, 582

\bibitem[{{Carnall} {et~al.}(2018){Carnall}, {McLure}, {Dunlop}, \& {Dav{\'e}}}]{Carnall2018MNRAS.480.4379C}
{Carnall}, A.~C., {McLure}, R.~J., {Dunlop}, J.~S., \& {Dav{\'e}}, R. 2018, \mnras, 480, 4379

\bibitem[{{Chen}(2012)}]{Chen2012MNRAS.427.1238C}
{Chen}, H.-W. 2012, \mnras, 427, 1238

\bibitem[{{Chen} {et~al.}(2010){Chen}, {Helsby}, {Gauthier}, {Shectman}, {Thompson}, \& {Tinker}}]{Chen2010ApJ...714.1521C}
{Chen}, H.-W., {Helsby}, J.~E., {Gauthier}, J.-R., {et~al.} 2010, \apj, 714, 1521

\bibitem[{{Chen} {et~al.}(2025){Chen}, {Wang}, {Zou}, {Zou}, {Gao}, {Wang}, {Yu}, {Jia}, {Li}, {Ma}, {Yao}, {Ding}, \& {Zhu}}]{Chen2025ApJ...981...81C}
{Chen}, Z., {Wang}, E., {Zou}, H., {et~al.} 2025, \apj, 981, 81

\bibitem[{{Churchill} {et~al.}(2013){Churchill}, {Nielsen}, {Kacprzak}, \& {Trujillo-Gomez}}]{Churchill2013ApJ...763L..42C}
{Churchill}, C.~W., {Nielsen}, N.~M., {Kacprzak}, G.~G., \& {Trujillo-Gomez}, S. 2013, \apjl, 763, L42

\bibitem[{{Cicone} {et~al.}(2016){Cicone}, {Maiolino}, \& {Marconi}}]{Cicone2016A&A...588A..41C}
{Cicone}, C., {Maiolino}, R., \& {Marconi}, A. 2016, \aap, 588, A41

\bibitem[{{Das} {et~al.}(2025){Das}, {Joshi}, {Chaudhary}, {Fumagalli}, {Fossati}, {P{\'e}roux}, \& {Ho}}]{Das2025A&A...695A.207D}
{Das}, S., {Joshi}, R., {Chaudhary}, R., {et~al.} 2025, \aap, 695, A207

\bibitem[{{Decataldo} {et~al.}(2024){Decataldo}, {Shen}, {Mayer}, {Baumschlager}, \& {Madau}}]{Decataldo2024A&A...685A...8D}
{Decataldo}, D., {Shen}, S., {Mayer}, L., {Baumschlager}, B., \& {Madau}, P. 2024, \aap, 685, A8

\bibitem[{{Dey} {et~al.}(2019){Dey}, {Schlegel}, {Lang}, {Blum}, {Burleigh}, {Fan}, {Findlay}, {Finkbeiner}, {Herrera}, {Juneau}, {Landriau}, {Levi}, {McGreer}, {Meisner}, {Myers}, {Moustakas}, {Nugent}, {Patej}, {Schlafly}, {Walker}, {Valdes}, {Weaver}, {Y{\`e}che}, {Zou}, {Zhou}, {Abareshi}, {Abbott}, {Abolfathi}, {Aguilera}, {Alam}, {Allen}, {Alvarez}, {Annis}, {Ansarinejad}, {Aubert}, {Beechert}, {Bell}, {BenZvi}, {Beutler}, {Bielby}, {Bolton}, {Brice{\~n}o}, {Buckley-Geer}, {Butler}, {Calamida}, {Carlberg}, {Carter}, {Casas}, {Castander}, {Choi}, {Comparat}, {Cukanovaite}, {Delubac}, {DeVries}, {Dey}, {Dhungana}, {Dickinson}, {Ding}, {Donaldson}, {Duan}, {Duckworth}, {Eftekharzadeh}, {Eisenstein}, {Etourneau}, {Fagrelius}, {Farihi}, {Fitzpatrick}, {Font-Ribera}, {Fulmer}, {G{\"a}nsicke}, {Gaztanaga}, {George}, {Gerdes}, {Gontcho}, {Gorgoni}, {Green}, {Guy}, {Harmer}, {Hernandez}, {Honscheid}, {Huang}, {James}, {Jannuzi}, {Jiang}, {Joyce}, {Karcher}, {Karkar}, {Kehoe}, {Kneib}, {Kueter-Young}, {Lan},
  {Lauer}, {Le Guillou}, {Le Van Suu}, {Lee}, {Lesser}, {Perreault Levasseur}, {Li}, {Mann}, {Marshall}, {Mart{\'\i}nez-V{\'a}zquez}, {Martini}, {du Mas des Bourboux}, {McManus}, {Meier}, {M{\'e}nard}, {Metcalfe}, {Mu{\~n}oz-Guti{\'e}rrez}, {Najita}, {Napier}, {Narayan}, {Newman}, {Nie}, {Nord}, {Norman}, {Olsen}, {Paat}, {Palanque-Delabrouille}, {Peng}, {Poppett}, {Poremba}, {Prakash}, {Rabinowitz}, {Raichoor}, {Rezaie}, {Robertson}, {Roe}, {Ross}, {Ross}, {Rudnick}, {Safonova}, {Saha}, {S{\'a}nchez}, {Savary}, {Schweiker}, {Scott}, {Seo}, {Shan}, {Silva}, {Slepian}, {Soto}, {Sprayberry}, {Staten}, {Stillman}, {Stupak}, {Summers}, {Sien Tie}, {Tirado}, {Vargas-Maga{\~n}a}, {Vivas}, {Wechsler}, {Williams}, {Yang}, {Yang}, {Yapici}, {Zaritsky}, {Zenteno}, {Zhang}, {Zhang}, {Zhou}, \& {Zhou}}]{Dey2019AJ....157..168D}
{Dey}, A., {Schlegel}, D.~J., {Lang}, D., {et~al.} 2019, \aj, 157, 168

\bibitem[{{Dutta} {et~al.}(2020){Dutta}, {Fumagalli}, {Fossati}, {Lofthouse}, {Prochaska}, {Arrigoni Battaia}, {Bielby}, {Cantalupo}, {Cooke}, {Murphy}, \& {O'Meara}}]{Dutta2020MNRAS.499.5022D}
{Dutta}, R., {Fumagalli}, M., {Fossati}, M., {et~al.} 2020, \mnras, 499, 5022

\bibitem[{{Elbaz} {et~al.}(2018){Elbaz}, {Leiton}, {Nagar}, {Okumura}, {Franco}, {Schreiber}, {Pannella}, {Wang}, {Dickinson}, {D{\'\i}az-Santos}, {Ciesla}, {Daddi}, {Bournaud}, {Magdis}, {Zhou}, \& {Rujopakarn}}]{Elbaz2018A&A...616A.110E}
{Elbaz}, D., {Leiton}, R., {Nagar}, N., {et~al.} 2018, \aap, 616, A110

\bibitem[{{Fern{\'a}ndez-Figueroa} {et~al.}(2024){Fern{\'a}ndez-Figueroa}, {Kacprzak}, {Nielsen}, {Barone}, {Nateghi}, {Sameer}, {Fisher}, \& {Chu}}]{Antonia2024MNRAS.531.3658F}
{Fern{\'a}ndez-Figueroa}, A., {Kacprzak}, G.~G., {Nielsen}, N.~M., {et~al.} 2024, \mnras, 531, 3658

\bibitem[{{Feroz} {et~al.}(2009){Feroz}, {Hobson}, \& {Bridges}}]{Feroz2009MNRAS.398.1601F}
{Feroz}, F., {Hobson}, M.~P., \& {Bridges}, M. 2009, \mnras, 398, 1601

\bibitem[{{Finkelstein} {et~al.}(2012){Finkelstein}, {Papovich}, {Salmon}, {Finlator}, {Dickinson}, {Ferguson}, {Giavalisco}, {Koekemoer}, {Reddy}, {Bassett}, {Conselice}, {Dunlop}, {Faber}, {Grogin}, {Hathi}, {Kocevski}, {Lai}, {Lee}, {McLure}, {Mobasher}, \& {Newman}}]{Finkelstein2012ApJ...756..164F}
{Finkelstein}, S.~L., {Papovich}, C., {Salmon}, B., {et~al.} 2012, \apj, 756, 164

\bibitem[{{Fossati} {et~al.}(2019){Fossati}, {Fumagalli}, {Gavazzi}, {Consolandi}, {Boselli}, {Yagi}, {Sun}, \& {Wilman}}]{Fossati2019MNRAS.484.2212F}
{Fossati}, M., {Fumagalli}, M., {Gavazzi}, G., {et~al.} 2019, \mnras, 484, 2212

\bibitem[{{Garcia} {et~al.}(2023){Garcia}, {Torrey}, {Hemler}, {Hernquist}, {Kewley}, {Nelson}, {Grasha}, {Zovaro}, \& {Chen}}]{Garcia2023MNRAS.519.4716G}
{Garcia}, A.~M., {Torrey}, P., {Hemler}, Z.~S., {et~al.} 2023, \mnras, 519, 4716

\bibitem[{{Gauthier}(2013)}]{Gauthier2013MNRAS.432.1444G}
{Gauthier}, J.-R. 2013, \mnras, 432, 1444

\bibitem[{{Girelli} {et~al.}(2020){Girelli}, {Pozzetti}, {Bolzonella}, {Giocoli}, {Marulli}, \& {Baldi}}]{Girelli2020A&A...634A.135G}
{Girelli}, G., {Pozzetti}, L., {Bolzonella}, M., {et~al.} 2020, \aap, 634, A135

\bibitem[{{Guo} {et~al.}(2023){Guo}, {Bacon}, {Bouch{\'e}}, {Wisotzki}, {Schaye}, {Blaizot}, {Verhamme}, {Cantalupo}, {Boogaard}, {Brinchmann}, {Cherrey}, {Kusakabe}, {Langan}, {Leclercq}, {Matthee}, {Michel-Dansac}, {Schroetter}, \& {Wendt}}]{Guo2023Natur.624...53G}
{Guo}, Y., {Bacon}, R., {Bouch{\'e}}, N.~F., {et~al.} 2023, \nat, 624, 53

\bibitem[{{Hafen} {et~al.}(2019){Hafen}, {Faucher-Gigu{\`e}re}, {Angl{\'e}s-Alc{\'a}zar}, {Stern}, {Kere{\v{s}}}, {Hummels}, {Esmerian}, {Garrison-Kimmel}, {El-Badry}, {Wetzel}, {Chan}, {Hopkins}, \& {Murray}}]{Hafen2019MNRAS.488.1248H}
{Hafen}, Z., {Faucher-Gigu{\`e}re}, C.-A., {Angl{\'e}s-Alc{\'a}zar}, D., {et~al.} 2019, \mnras, 488, 1248

\bibitem[{{Ho} {et~al.}(2017){Ho}, {Martin}, {Kacprzak}, \& {Churchill}}]{Ho2017ApJ...835..267H}
{Ho}, S.~H., {Martin}, C.~L., {Kacprzak}, G.~G., \& {Churchill}, C.~W. 2017, \apj, 835, 267

\bibitem[{{Ho} {et~al.}(2020){Ho}, {Martin}, \& {Schaye}}]{Ho2020ApJ...904...76H}
{Ho}, S.~H., {Martin}, C.~L., \& {Schaye}, J. 2020, \apj, 904, 76

\bibitem[{{Ho} {et~al.}(2021){Ho}, {Martin}, \& {Schaye}}]{Ho2021ApJ...923..137H}
{Ho}, S.~H., {Martin}, C.~L., \& {Schaye}, J. 2021, \apj, 923, 137

\bibitem[{{Hopkins} {et~al.}(2012){Hopkins}, {Quataert}, \& {Murray}}]{Hopkins2012MNRAS.421.3522H}
{Hopkins}, P.~F., {Quataert}, E., \& {Murray}, N. 2012, \mnras, 421, 3522

\bibitem[{{Huang} {et~al.}(2019){Huang}, {Zou}, {Kong}, {Comparat}, {Lin}, {Gao}, {Liang}, {Delubac}, {Raichoor}, {Kneib}, {Schneider}, {Zhou}, {Yuan}, \& {Bershady}}]{Huang2019ApJ...886...31H}
{Huang}, C., {Zou}, H., {Kong}, X., {et~al.} 2019, \apj, 886, 31

\bibitem[{{Huang} {et~al.}(2021){Huang}, {Chen}, {Shectman}, {Johnson}, {Zahedy}, {Helsby}, {Gauthier}, \& {Thompson}}]{Huang2021MNRAS.502.4743H}
{Huang}, Y.-H., {Chen}, H.-W., {Shectman}, S.~A., {et~al.} 2021, \mnras, 502, 4743

\bibitem[{{Huscher} {et~al.}(2021){Huscher}, {Oppenheimer}, {Lonardi}, {Crain}, {Richings}, \& {Schaye}}]{Huscher2021MNRAS.500.1476H}
{Huscher}, E., {Oppenheimer}, B.~D., {Lonardi}, A., {et~al.} 2021, \mnras, 500, 1476

\bibitem[{{Joshi} {et~al.}(2025){Joshi}, {Das}, {Fumagalli}, {Fossati}, {P{\'e}roux}, {Chaudhary}, {Yesuf}, \& {Ho}}]{Joshi2025A&A...695A.206J}
{Joshi}, R., {Das}, S., {Fumagalli}, M., {et~al.} 2025, \aap, 695, A206

\bibitem[{{Joshi} {et~al.}(2017){Joshi}, {Srianand}, {Petitjean}, \& {Noterdaeme}}]{Joshi2017MNRAS.471.1910J}
{Joshi}, R., {Srianand}, R., {Petitjean}, P., \& {Noterdaeme}, P. 2017, \mnras, 471, 1910

\bibitem[{{Kacprzak} {et~al.}(2008){Kacprzak}, {Churchill}, {Steidel}, \& {Murphy}}]{Kacprzak2008AJ....135..922K}
{Kacprzak}, G.~G., {Churchill}, C.~W., {Steidel}, C.~C., \& {Murphy}, M.~T. 2008, \aj, 135, 922

\bibitem[{{Kacprzak} {et~al.}(2010){Kacprzak}, {Murphy}, \& {Churchill}}]{Kacprzak2010MNRAS.406..445K}
{Kacprzak}, G.~G., {Murphy}, M.~T., \& {Churchill}, C.~W. 2010, \mnras, 406, 445

\bibitem[{{Kacprzak} {et~al.}(2025){Kacprzak}, {Oppenheimer}, {Nielsen}, {Fernandez-Figueroa}, {Murphy}, {Allen}, {Barone}, {Sameer}, {Churchill}, {Burchett}, {Gupta}, {Charlton}, \& {Platukis}}]{Kacprzak2025arXiv250711613K}
{Kacprzak}, G.~G., {Oppenheimer}, B.~D., {Nielsen}, N.~M., {et~al.} 2025, arXiv e-prints, arXiv:2507.11613

\bibitem[{{Lan} \& {Mo}(2018)}]{Lan2018ApJ...866...36L}
{Lan}, T.-W. \& {Mo}, H. 2018, \apj, 866, 36

\bibitem[{{Lang} {et~al.}(2016){Lang}, {Hogg}, \& {Schlegel}}]{Lang2016AJ....151...36L}
{Lang}, D., {Hogg}, D.~W., \& {Schlegel}, D.~J. 2016, \aj, 151, 36

\bibitem[{{Langeroodi} \& {Hjorth}(2024)}]{Langeroodi2024arXiv240907455L}
{Langeroodi}, D. \& {Hjorth}, J. 2024, arXiv e-prints, arXiv:2409.07455

\bibitem[{{Lequeux} {et~al.}(1979){Lequeux}, {Peimbert}, {Rayo}, {Serrano}, \& {Torres-Peimbert}}]{LequeuxJ.1979A&A....80..155L}
{Lequeux}, J., {Peimbert}, M., {Rayo}, J.~F., {Serrano}, A., \& {Torres-Peimbert}, S. 1979, \aap, 80, 155

\bibitem[{{L{\'o}pez} \& {Chen}(2012)}]{Lopez2012MNRAS.419.3553L}
{L{\'o}pez}, G. \& {Chen}, H.-W. 2012, \mnras, 419, 3553

\bibitem[{{Lundgren} {et~al.}(2021){Lundgren}, {Creech}, {Brammer}, {Kirse}, {Peek}, {Wake}, {York}, {Chisholm}, {Erb}, {Kulkarni}, {Straka}, {Tremonti}, \& {van Dokkum}}]{Lundgren2021ApJ...913...50L}
{Lundgren}, B.~F., {Creech}, S., {Brammer}, G., {et~al.} 2021, \apj, 913, 50

\bibitem[{{Mannucci} {et~al.}(2010){Mannucci}, {Cresci}, {Maiolino}, {Marconi}, \& {Gnerucci}}]{MannucciF.2010MNRAS.408.2115M}
{Mannucci}, F., {Cresci}, G., {Maiolino}, R., {Marconi}, A., \& {Gnerucci}, A. 2010, \mnras, 408, 2115

\bibitem[{{Murray} {et~al.}(2011){Murray}, {M{\'e}nard}, \& {Thompson}}]{Murray2011ApJ...735...66M}
{Murray}, N., {M{\'e}nard}, B., \& {Thompson}, T.~A. 2011, \apj, 735, 66

\bibitem[{{Napolitano} {et~al.}(2023){Napolitano}, {Pandey}, {Myers}, {Lan}, {Anand}, {Aguilar}, {Ahlen}, {Alexander}, {Brooks}, {Canning}, {Circosta}, {De La Macorra}, {Doel}, {Eftekharzadeh}, {Fawcett}, {Font-Ribera}, {Garcia-Bellido}, {Gontcho A Gontcho}, {Le Guillou}, {Guy}, {Honscheid}, {Juneau}, {Kisner}, {Landriau}, {Meisner}, {Miquel}, {Moustakas}, {Percival}, {Prochaska}, {Schubnell}, {Tarl{\'e}}, {Weaver}, {Weiner}, {Zhou}, {Zou}, \& {Zou}}]{Napolitano2023AJ....166...99N}
{Napolitano}, L., {Pandey}, A., {Myers}, A.~D., {et~al.} 2023, \aj, 166, 99

\bibitem[{{Nateghi} {et~al.}(2024){Nateghi}, {Kacprzak}, {Nielsen}, {Sameer}, {Murphy}, {Churchill}, \& {Charlton}}]{Nateghi2024MNRAS.534..930N}
{Nateghi}, H., {Kacprzak}, G.~G., {Nielsen}, N.~M., {et~al.} 2024, \mnras, 534, 930

\bibitem[{{Nelson} {et~al.}(2020){Nelson}, {Sharma}, {Pillepich}, {Springel}, {Pakmor}, {Weinberger}, {Vogelsberger}, {Marinacci}, \& {Hernquist}}]{Nelson2020MNRAS.498.2391N}
{Nelson}, D., {Sharma}, P., {Pillepich}, A., {et~al.} 2020, \mnras, 498, 2391

\bibitem[{{Nestor} {et~al.}(2011){Nestor}, {Johnson}, {Wild}, {M{\'e}nard}, {Turnshek}, {Rao}, \& {Pettini}}]{Nestor2011MNRAS.412.1559N}
{Nestor}, D.~B., {Johnson}, B.~D., {Wild}, V., {et~al.} 2011, \mnras, 412, 1559

\bibitem[{{Nielsen} {et~al.}(2013){Nielsen}, {Churchill}, \& {Kacprzak}}]{Nielsen2013ApJ...776..115N}
{Nielsen}, N.~M., {Churchill}, C.~W., \& {Kacprzak}, G.~G. 2013, \apj, 776, 115

\bibitem[{{Nielsen} {et~al.}(2018){Nielsen}, {Kacprzak}, {Pointon}, {Churchill}, \& {Murphy}}]{Nielsen2018ApJ...869..153N}
{Nielsen}, N.~M., {Kacprzak}, G.~G., {Pointon}, S.~K., {Churchill}, C.~W., \& {Murphy}, M.~T. 2018, \apj, 869, 153

\bibitem[{{Nielsen} {et~al.}(2020){Nielsen}, {Kacprzak}, {Pointon}, {Murphy}, {Churchill}, \& {Dav{\'e}}}]{Nielsen2020ApJ...904..164N}
{Nielsen}, N.~M., {Kacprzak}, G.~G., {Pointon}, S.~K., {et~al.} 2020, \apj, 904, 164

\bibitem[{{Noterdaeme} {et~al.}(2010){Noterdaeme}, {Srianand}, \& {Mohan}}]{Noterdaeme2010MNRAS.403..906N}
{Noterdaeme}, P., {Srianand}, R., \& {Mohan}, V. 2010, \mnras, 403, 906

\bibitem[{{Peroux} \& {Nelson}(2024)}]{Peroux2024arXiv241107988P}
{Peroux}, C. \& {Nelson}, D. 2024, arXiv e-prints, arXiv:2411.07988

\bibitem[{{P{\'e}roux} {et~al.}(2020){P{\'e}roux}, {Nelson}, {van de Voort}, {Pillepich}, {Marinacci}, {Vogelsberger}, \& {Hernquist}}]{Peroux2020MNRAS.499.2462P}
{P{\'e}roux}, C., {Nelson}, D., {van de Voort}, F., {et~al.} 2020, \mnras, 499, 2462

\bibitem[{{P{\'e}roux} {et~al.}(2017){P{\'e}roux}, {Rahmani}, {Quiret}, {Pettini}, {Kulkarni}, {York}, {Straka}, {Husemann}, \& {Milliard}}]{Peroux2017MNRAS.464.2053P}
{P{\'e}roux}, C., {Rahmani}, H., {Quiret}, S., {et~al.} 2017, \mnras, 464, 2053

\bibitem[{{Ponti} {et~al.}(2023){Ponti}, {Sanders}, {Locatelli}, {Zheng}, {Zhang}, {Yeung}, {Freyberg}, {Dennerl}, {Comparat}, {Merloni}, {Di Teodoro}, {Sasaki}, \& {Reiprich}}]{Ponti2023A&A...670A..99P}
{Ponti}, G., {Sanders}, J.~S., {Locatelli}, N., {et~al.} 2023, \aap, 670, A99

\bibitem[{{Popesso} {et~al.}(2023){Popesso}, {Concas}, {Cresci}, {Belli}, {Rodighiero}, {Inami}, {Dickinson}, {Ilbert}, {Pannella}, \& {Elbaz}}]{Popesso2023MNRAS.519.1526P}
{Popesso}, P., {Concas}, A., {Cresci}, G., {et~al.} 2023, \mnras, 519, 1526

\bibitem[{{Rahmati} {et~al.}(2015){Rahmati}, {Schaye}, {Bower}, {Crain}, {Furlong}, {Schaller}, \& {Theuns}}]{Rahmati2015MNRAS.452.2034R}
{Rahmati}, A., {Schaye}, J., {Bower}, R.~G., {et~al.} 2015, \mnras, 452, 2034

\bibitem[{{Ramesh} \& {Nelson}(2024)}]{Ramesh2024MNRAS.528.3320R}
{Ramesh}, R. \& {Nelson}, D. 2024, \mnras, 528, 3320

\bibitem[{{Reichardt Chu} {et~al.}(2025){Reichardt Chu}, {Fisher}, {Chisholm}, {Berg}, {Bolatto}, {Cameron}, {Fielding}, {Herrera-Camus}, {Kacprzak}, {Li}, {McLeod}, {McPherson}, {Nielsen}, {Rickards Vaught}, {Ridolfo}, \& {Sandstrom}}]{Chu2025MNRAS.536.1799R}
{Reichardt Chu}, B., {Fisher}, D.~B., {Chisholm}, J., {et~al.} 2025, \mnras, 536, 1799

\bibitem[{{Rubin} {et~al.}(2018){Rubin}, {Diamond-Stanic}, {Coil}, {Crighton}, \& {Moustakas}}]{Rubin2018ApJ...853...95R}
{Rubin}, K. H.~R., {Diamond-Stanic}, A.~M., {Coil}, A.~L., {Crighton}, N. H.~M., \& {Moustakas}, J. 2018, \apj, 853, 95

\bibitem[{{Rudie} {et~al.}(2019){Rudie}, {Steidel}, {Pettini}, {Trainor}, {Strom}, {Hummels}, {Reddy}, \& {Shapley}}]{Rudie2019ApJ...885...61R}
{Rudie}, G.~C., {Steidel}, C.~C., {Pettini}, M., {et~al.} 2019, \apj, 885, 61

\bibitem[{{Schroetter} {et~al.}(2019){Schroetter}, {Bouch{\'e}}, {Zabl}, {Contini}, {Wendt}, {Schaye}, {Mitchell}, {Muzahid}, {Marino}, {Bacon}, {Lilly}, {Richard}, \& {Wisotzki}}]{Schroetter2019MNRAS.490.4368S}
{Schroetter}, I., {Bouch{\'e}}, N.~F., {Zabl}, J., {et~al.} 2019, \mnras, 490, 4368

\bibitem[{{Schroetter} {et~al.}(2024){Schroetter}, {Bouch{\'e}}, {Zabl}, {Wendt}, {Cherrey}, {Langan}, {Schaye}, \& {Contini}}]{Schroetter2024A&A...687A..39S}
{Schroetter}, I., {Bouch{\'e}}, N.~F., {Zabl}, J., {et~al.} 2024, \aap, 687, A39

\bibitem[{{Straka} {et~al.}(2015){Straka}, {Noterdaeme}, {Srianand}, {Nutalaya}, {Kulkarni}, {Khare}, {Bowen}, {Bishof}, \& {York}}]{Straka2015MNRAS.447.3856S}
{Straka}, L.~A., {Noterdaeme}, P., {Srianand}, R., {et~al.} 2015, \mnras, 447, 3856

\bibitem[{{Suresh} {et~al.}(2019){Suresh}, {Nelson}, {Genel}, {Rubin}, \& {Hernquist}}]{Suresh2019MNRAS.483.4040S}
{Suresh}, J., {Nelson}, D., {Genel}, S., {Rubin}, K. H.~R., \& {Hernquist}, L. 2019, \mnras, 483, 4040

\bibitem[{{Tremonti} {et~al.}(2004){Tremonti}, {Heckman}, {Kauffmann}, {Brinchmann}, {Charlot}, {White}, {Seibert}, {Peng}, {Schlegel}, {Uomoto}, {Fukugita}, \& {Brinkmann}}]{Tremonti2004ApJ...613..898T}
{Tremonti}, C.~A., {Heckman}, T.~M., {Kauffmann}, G., {et~al.} 2004, \apj, 613, 898

\bibitem[{{Tumlinson} {et~al.}(2017){Tumlinson}, {Peeples}, \& {Werk}}]{Tumlinson2017ARA&A..55..389T}
{Tumlinson}, J., {Peeples}, M.~S., \& {Werk}, J.~K. 2017, \araa, 55, 389

\bibitem[{Weng {et~al.}(2024)Weng, P{\'e}roux, Ramesh, Nelson, Sadler, Zwaan, Bollo, \& Casavecchia}]{weng2024physical}
Weng, S., P{\'e}roux, C., Ramesh, R., {et~al.} 2024, Monthly Notices of the Royal Astronomical Society, 527, 3494

\bibitem[{{Weng} {et~al.}(2024){Weng}, {P{\'e}roux}, {Ramesh}, {Nelson}, {Sadler}, {Zwaan}, {Bollo}, \& {Casavecchia}}]{Weng2024MNRAS.527.3494W}
{Weng}, S., {P{\'e}roux}, C., {Ramesh}, R., {et~al.} 2024, \mnras, 527, 3494

\bibitem[{{Wu} {et~al.}(2025){Wu}, {Cai}, {Lan}, {Zou}, {Anand}, {Dey}, {Li}, {Aguilar}, {Ahlen}, {Brooks}, {Claybaugh}, {de la Macorra}, {Doel}, {Ferraro}, {Forero-Romero}, {Gontcho A Gontcho}, {Honscheid}, {Juneau}, {Kehoe}, {Kisner}, {Lambert}, {Landriau}, {Le Guillou}, {Manera}, {Meisner}, {Miquel}, {Moustakas}, {Newman}, {Prada}, {Rossi}, {Sanchez}, {Schlegel}, {Schubnell}, {Siudek}, {Sprayberry}, {Tarl{\'e}}, {Weaver}, \& {Zou}}]{Wu2025ApJ...983..186W}
{Wu}, X., {Cai}, Z., {Lan}, T.~W., {et~al.} 2025, \apj, 983, 186

\bibitem[{{Zabl} {et~al.}(2019){Zabl}, {Bouch{\'e}}, {Schroetter}, {Wendt}, {Finley}, {Schaye}, {Conseil}, {Contini}, {Marino}, {Mitchell}, {Muzahid}, {Pezzulli}, \& {Wisotzki}}]{Zabl2019MNRAS.485.1961Z}
{Zabl}, J., {Bouch{\'e}}, N.~F., {Schroetter}, I., {et~al.} 2019, \mnras, 485, 1961

\bibitem[{{Zabl} {et~al.}(2021){Zabl}, {Bouch{\'e}}, {Wisotzki}, {Schaye}, {Leclercq}, {Garel}, {Wendt}, {Schroetter}, {Muzahid}, {Cantalupo}, {Contini}, {Bacon}, {Brinchmann}, \& {Richard}}]{Zabl2021MNRAS.507.4294Z}
{Zabl}, J., {Bouch{\'e}}, N.~F., {Wisotzki}, L., {et~al.} 2021, \mnras, 507, 4294

\bibitem[{{Zahedy} {et~al.}(2019){Zahedy}, {Chen}, {Johnson}, {Pierce}, {Rauch}, {Huang}, {Weiner}, \& {Gauthier}}]{Zahedy2019MNRAS.484.2257Z}
{Zahedy}, F.~S., {Chen}, H.-W., {Johnson}, S.~D., {et~al.} 2019, \mnras, 484, 2257

\bibitem[{{Zahid} {et~al.}(2014){Zahid}, {Dima}, {Kudritzki}, {Kewley}, {Geller}, {Hwang}, {Silverman}, \& {Kashino}}]{Zahid2014ApJ...791..130Z}
{Zahid}, H.~J., {Dima}, G.~I., {Kudritzki}, R.-P., {et~al.} 2014, \apj, 791, 130

\bibitem[{{Zahid} {et~al.}(2011){Zahid}, {Kewley}, \& {Bresolin}}]{Zahid2011ApJ...730..137Z}
{Zahid}, H.~J., {Kewley}, L.~J., \& {Bresolin}, F. 2011, \apj, 730, 137

\bibitem[{{Zhu} \& {M{\'e}nard}(2013)}]{Zhu2013ApJ...770..130Z}
{Zhu}, G. \& {M{\'e}nard}, B. 2013, \apj, 770, 130

\bibitem[{{Zhuang} \& {Ho}(2019)}]{Zhuang2019ApJ...882...89Z}
{Zhuang}, M.-Y. \& {Ho}, L.~C. 2019, \apj, 882, 89

\bibitem[{{Zibetti} {et~al.}(2007){Zibetti}, {M{\'e}nard}, {Nestor}, {Quider}, {Rao}, \& {Turnshek}}]{Zibetti2007ApJ...658..161Z}
{Zibetti}, S., {M{\'e}nard}, B., {Nestor}, D.~B., {et~al.} 2007, \apj, 658, 161

\end{thebibliography}
\end{document}